\documentclass[10pt,letterpaper]{article}
\usepackage[top=0.85in,left=1.25in,footskip=0.75in]{geometry}

\usepackage{amsmath,amssymb}

\usepackage{changepage}

\usepackage[utf8x]{inputenc}

\usepackage{textcomp,marvosym}

\usepackage{cite}

\usepackage{nameref,hyperref}

\usepackage{microtype}
\DisableLigatures[f]{encoding = *, family = * }

\usepackage[table]{xcolor}

\usepackage{array}

\newcolumntype{+}{!{\vrule width 2pt}}

\newlength\savedwidth

\newcommand\thickhline{\noalign{\global\savedwidth\arrayrulewidth\global\arrayrulewidth 2pt}%
\hline
\noalign{\global\arrayrulewidth\savedwidth}}

\usepackage[aboveskip=1pt,labelfont=bf,labelsep=period,justification=raggedright,singlelinecheck=off]{caption}

\makeatletter
\renewcommand{\@biblabel}[1]{\quad#1.}
\makeatother

\usepackage{lastpage,fancyhdr,graphicx}
\usepackage{epstopdf}
\usepackage{color}
\usepackage{multirow}

\newcommand{\pan}[1]{\left\langle#1\right\rangle}
\newcommand{\udef}{\stackrel{def}{=}}

\newcommand{\compart}[1]{\ulcorner \mathbf{#1} \lrcorner}
\newcommand{\dfs}{{\bfseries dfs}}
\newcommand{\wt}{{\itshape wt}}
\newcommand{\dfsm}{\mathit{dfs}}
\newcommand{\dfsi}{\textit{dfs}}
\newcommand{\sbab}{\blacksquare}

\begin{document}
\vspace*{0.2in}

\begin{flushleft}
{\Large
\textbf{Endemicity and prevalence of multipartite viruses under heterogeneous between-host transmission}
}
\newline
\\
{\large
Eugenio Valdano\textsuperscript{1$\ast\dagger$},
Susanna Manrubia\textsuperscript{2,3},
Sergio G\'{o}mez\textsuperscript{1},
Alex Arenas\textsuperscript{1}
}
\\
\bigskip
\textbf{1} Departament d'Enginyeria Inform\`{a}tica i Matem\`{a}tiques, Universitat Rovira i Virgili, 43007 Tarragona, Spain.
\\
\textbf{2} National Centre for Biotechnology (CSIC), Madrid, Spain
\\
\textbf{3} Grupo Interdisciplinar de Sistemas Complejos (GISC), Madrid, Spain
\\
\bigskip

$\ast$ \href{mailto:eugenio.valdano@gmail.com}{eugenio.valdano@gmail.com} \\
$\dagger$ Current affiliation: Center for Biomedical Modeling,
The Semel Institute for Neuroscience and Human Behavior,
David Geffen School of Medicine,
University of California Los Angeles,
Los Angeles,
CA 90024, USA.

\end{flushleft}
\section*{Abstract}
Multipartite viruses replicate through a puzzling evolutionary strategy. Their genome is segmented into two or more parts, and encapsidated in separate particles that appear to propagate independently. Completing the replication cycle, however, requires the full genome, so that a systemic infection of a host requires the concurrent presence of several particles. This represents an apparent evolutionary drawback of multipartitism, while its advantages remain unclear.
A transition from monopartite to multipartite viral forms has been described { \it in vitro} under conditions of high multiplicity of infection, suggesting that cooperation between defective mutants is a plausible evolutionary pathway towards multipartitism. However, it is unknown how the putative advantages that multipartitism might enjoy at the microscopic level affect its epidemiology, or if an explicit advantange is needed to explain its ecological persistence. In order to disentangle which mechanisms might contribute to the rise and fixation of multipartitism, we here investigate the interaction between viral spreading dynamics and host population structure. We set up a compartmental model of the spread of a virus in its different forms and explore its epidemiology using both analytical and numerical techniques. We uncover that the impact of host contact structure on spreading dynamics entails a rich phenomenology of ecological relationships that includes cooperation, competition, and commensality. Furthermore, we find out that multipartitism might rise to fixation even in the absence of explicit microscopic advantages. Multipartitism allows the virus to colonize environments that could not be invaded by the monopartite form, while homogeneous contacts between hosts facilitate its spread. We conjecture that these features might have led to an increase in the diversity and prevalence of multipartite viral forms concomitantly with the expansion of agricultural practices.

\section*{Author summary}
Viruses typically consist of some genetic material wrapped up in a single particle, the capsid. Multipartite viruses follow another lifestyle. Their genome is made up of several segments, each packed in independent particles. However, since the completion of the viral cycle requires the full genome, these particles need to coinfect each host. This imposes strong constraints on the minimum number of independently transmitted particles, making the rise and persistence of multipartitism an evolutionary puzzle. By using analytical and numerical tools, we study the ecological interaction between monopartite and multipartite forms, in terms of their ability to spread on, and take over, a host population. We reveal that this interaction can take various forms (competition, cooperation, commensality), depending on the underlying structure of contacts among hosts. We also find that, in some situations, multipartitism represents an effective adaptive strategy, allowing the virus to colonize environments in which the monopartite form cannot thrive. Finally, we uncover that contact structures typical of farmed plants favor multipartitism, suggesting a correlation between the intensification of agricultural practices and an increase in the diversity and prevalence of multipartite viral species.


\section*{Introduction}

Viruses transport their genetic material inside a protein shell, the capsid, surrounded in some species by a lipid membrane. In most viral species, each viral particle contains all the genetic material needed to carry out replication inside a host cell, and generate a progeny of viral particles. A prominent exception to this behavior is found in multipartite viruses. These viruses, first described in the 1960s~\cite{lister:1966}, have a genome segmented in two or more parts. According to current evidence, the segments are encapsidated separately and, apparently, propagate independently~\cite{Sicard2016,dallara:2016}.
As of today, there is no mainstream theory able to explain the adaptive advantage of such a strange lifestyle~\cite{wu:2017}. The main puzzle regarding multipartite viruses is how the simultaneous presence of multiple segments, which imposes severe constraints on the number of viral particles that have to reach a susceptible host, is balanced by other adaptive advantages of multipartitism, whether microscopic or ecological~\cite{Sanz2017,lucia2018theoretical}.

Despite this apparent paradox, multipartitism is widespread in the Virosphere, as up to 40\% of all known viral families are multipartite \cite{hull2013plant}.
A large majority of them infects either plants or fungi, with only four known examples of species infecting exclusively animals~\cite{Sanz2017}. Evolutionary pathways leading to multipartitism are likely to be multiple, since this strategy is present in RNA and DNA viruses, and in the latter case an origin to a single ancestral virus cannot be traced. Beyond its virological interest, multipartitism has a particularly negative effect on agricultural production, as several multipartite viruses are pathogenic, and routinely cripple crop yield~\cite{moreno:2008}. Cultivars themselves may have directly played a role in the rise of multipartitism, as an evolutionary radiation in the diversity of viral species, many being just centuries old, was likely promoted by an intensification of agricultural practices~\cite{fargette2008diversification,pagan2010long,gibbs2008prehistory}. It has been put forward that multipartite species might be at an advantage in the face of environmental changes, since they likely adapt faster due to new combinations of segments promoted by their genomic architecture~\cite{Sanz2017}. It is known that changes in land cover offer multiple opportunities for novel interactions between plants and pathogens \cite{burdon2006current,jones2009plant,pagan2012effect,alexander2014plant}. Studies on the impact of agriculture in viral ecology have uncovered a surprisingly negative association between plant diversity and family-level diversity of plant-associated viruses, and a higher prevalence of viruses in cultivated areas \cite{bernardo2017geometagenomics}.

The emergence of defective variants out of the wild-type (\wt{}) form (i.e., the one containing the full genome~\cite{huang:1973}) has been both posited and observed in controlled environments, arising from replication errors and thriving under conditions that ensure high multiplicity of infection (MOI)~\cite{perrault:1981,schlesinger:1988}. Specifically, it has been shown {\it in vitro} that two defective forms spontaneously generated by foot-and-mouth disease virus (FMDV has an unsegmented genome formed by ssRNA of positive polarity) can complement each other and quickly substitute the \wt{} form~\cite{garcia-arriaza:2004}. This strategy has been formally explored in models of competitive dynamics between the \wt{} and a number of complementing segments~\cite{Iranzo2012}, which implemented different advantages that could compensate for the cost of an increased MOI. The model mimicked the experimental setting where, in particular, host-cell availability corresponded to that of a well-mixed system. Two of the advantages implemented had been theoretically proposed in the past, though, as of yet, have not received empirical support (a faster replicative ability~\cite{chao:1991} and a slower accumulation of deleterious mutations in shorter segments~\cite{pressing:1984,nee:1990}) while, in the case of FMDV, it was shown that capsids containing shorter genomes enjoyed a larger average lifetime between infection events~\cite{ojosnegros:2011}. This differential degradation, dependent on genome length, was sufficient to compensate for co-infection requirements in multipartite forms with two, to up to four, segments~\cite{Iranzo2012}, but cannot explain the emergence of multipartite viruses with many segments, such as nanoviruses or babuviruses \cite{Sanz2017}. Hence, the evolutionary pathway explored in that work would be applicable to a subset of all currently described multipartite viral species.

What is missing from this picture is investigating how the interaction between viral dynamics and host ecology shapes the rise and persistence of multipartite viral forms at the host population level.
We also wish to quantify the impact of different host contact structures in driving the success of multipartitism.
We tackle this problem by building a compartmental model for studying the competition between monopartite and multipartite variants in terms of their ability to spread and persist on a structured host population.

As the generation of functional defective mutants from the \wt{} occurs at a much longer time scale than the spread of the virus in the population, we set up a model that already contains both the \wt{} and a cohort of defective forms, potentially complementary. This allows us to study the competition dynamics between the different forms causing them to coexist in, or take over the ecological niche. Using both analytical calculations and numerical tools, we investigate the outcome of a random emergence of mutants, and derive the conditions that make multipartitism a fitness-enhancing strategy, allowing the virus to adapt to a wider range of hosts and environments. Since no apparent structural feature discriminates multipartite viruses from monopartite ones --~they are found exhibiting different capsid structures, genome sizes and types~\cite{Sicard2015}~-- we include in the model as few virological features as possible, and investigate how multipartitism impacts on the spreading potential of the virus.
We do, however, account for key viral mechanisms that can drive the resilience of multipartitism in an ecological context. The first one is the already mentioned differential degradation, i.e., the different average lifetime of defective viral particles with respect to \wt{}'s, or formally equivalent advantages of faster replication or elimination of deleterious mutations through sex. A second biological mechanism is the mode of transmission of multipartite viruses between hosts. Most known multipartite viruses are spread by vectors (mainly insects), which typically pick up very few viral particles from an infected plant~\cite{Moury2007}. The transmission process between hosts typically acts as a population bottleneck for the virus, entailing a loss of genetic diversity and, if severe enough, the systematic purge of deleterious forms~\cite{Manrubia2010,Gallet2017}. Thanks to our parsimonious modeling setup, any of the aforementioned mechanisms can be seen as effectively impacting the chances of the \wt{} or defective particles to reach the target hosts, leading to a difference in transmissibility.
A single model parameter, therefore, by tuning this relative transmissibility, embraces a number of different biological processes. In this sense, the results of our model can be extended to other systems as long as the specific mechanisms involved in their spreading fitness can be cast in the form of changes in transmissibility. Remarkably, we find out that even in the absence of an explicit microscopic advantage, ecological dynamics might cause the fixation of the multipartite form due to the stochastic extinction (analogous to random drift) of the monopartite virus.

Alongside the biological properties of the virus, the model implements the structure of contacts among susceptible hosts through which the  viruses can spread. Often, in our context, this means the contact network induced by vector movements among plants. Its topology may be diverse, depending on plant distribution and vector behavior, with two limit cases being the distribution of plant species in the wild~(see, e.g.\cite{denny:2017} and references therein) and huge modern agriculturally homogeneous regions~\cite{leff:2004}. These different architectures are implemented by tuning the distribution of contact rates among hosts, with limiting cases being fixed contact rate, and power law-distributed contact rate. This feature is the key tool to uncover how the structure of contacts among hosts shapes the endemicity and prevalence of the different viral forms.

We remark that compartmental models of interacting diseases have been studied in the past~\cite{zhang2001synergism,Rohani2003,Newman2005,AbuRaddad2006,Karrer2011,Poletto2013,perefarres2014frequency}. Those models, however, assume that the disease agents involved are fully-fledged pathogens that can spread on their own, and cannot describe asymmetric viral associations~\cite{sofonea2017exposing}, of which multipartitism is an example. Thanks to that, a completely new phenomenology emerges, driven by a complex evolutionary dynamics, and involving a wide range of ecological interactions: competition, symbiosis, commensality.

\section*{Results}

\subsection*{The model}

We consider a large population of $N$ hosts susceptible to a virus that may circulate in its {\itshape wild-type} ($wt$) form together with $v$ defective variants that are potentially complementing, i.e., when simultaneously present in the host, they are able to complete the virus infective cycle even in the absence of the \wt{}. As previously done~\cite{Iranzo2012}, we take advantage of the timescale separation between the random emergence of defective mutants due to errors during replication of the complete genome, and the competition between forms driven by their spread in the host population. This allows us to effectively model the random emergence of mutants as an initial small, yet nonzero, prevalence in the host population.

\begin{figure}[htbp]
	\includegraphics[width=\textwidth]{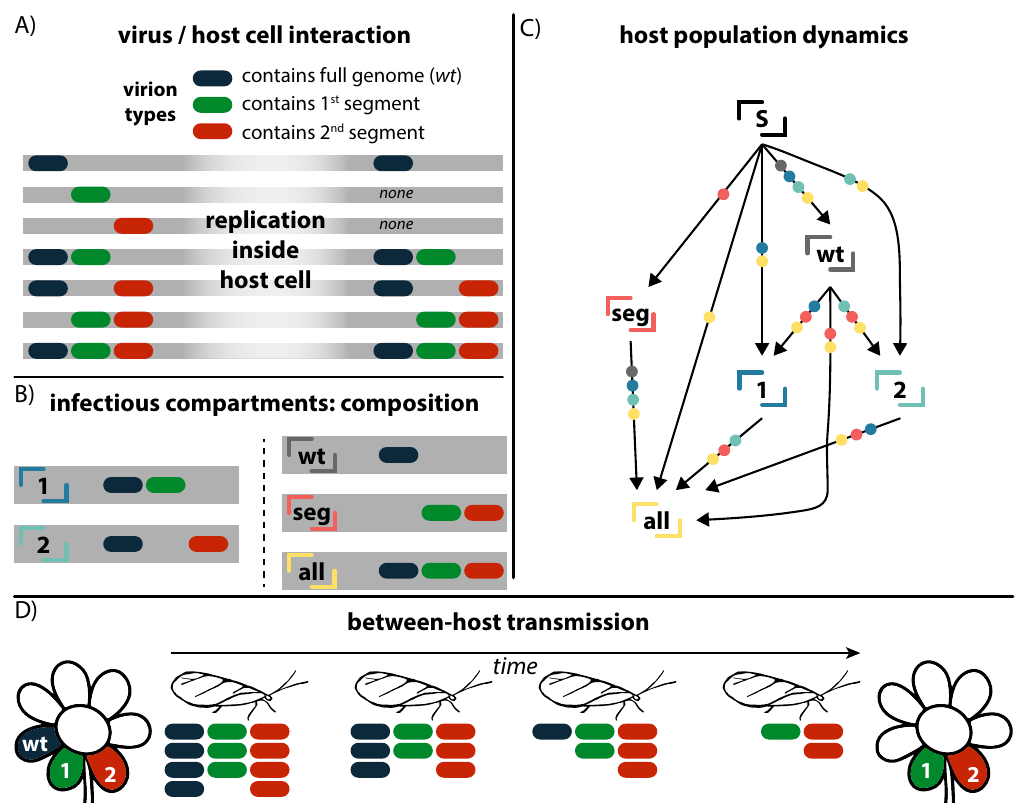} 
	\caption{Schematic illustration of viral dynamics and modelling framework in the case of a bipartite virus.
		{\itshape A)} describes the different viral species circulating, and their replication dynamics inside a host cell. In each line, the viral particles infecting the same host cell are shown on the left, and the product of replication on the right.
		{\itshape B)} describes the different infectious compartments of the model at population level, in terms of the viral species they are infected by.
		In {\itshape C)} we outline the compartmental model of a bipartite virus. An arrow going from one compartment to another means that a host in the former state can move to the latter by coming into contact with one of the compartments marked as dots on the arrow itself. Here, we show neither the recovery rate ($\mu$) at which infectious compartments turn susceptible, nor the transmission rates corresponding to each interaction.
		{\itshape D)} illustrates the vector-mediated viral transmission from host to host. The vector picks up some viral particles of different variants (represented in the figure below the vector itself). During the time it takes for it to reach another plant, these particles degrade. One hypothesis behind differential transmissibility is differential degradation, here depicted. Lower degradation rates due of the defective variants lead to a chance of transmission higher than \wt{}.}
	\label{fig:illustration}
\end{figure}

As customary in compartmental models, we consider that hosts are either free of the virus, and thus susceptible (S), or infected by a certain combination of the viral forms, translating into various infectious compartments (Fig.~\ref{fig:illustration}). The main assumption is that a host can be infected only by a combination that guarantees the presence of the full genome. Without it, there is no completion of the viral cycle, and thus no systemic infection is possible. Moreover, we assume that host cells replicate all, and only, the viral forms they are infected by (Fig.~\ref{fig:illustration}A). These assumptions determine the set of existing compartments.

Two infectious compartments are present regardless of the value of $v$. They are $\compart{wt}$ and $\compart{all}$, and correspond to plants infected by the \wt{} only, and by the \wt{} together with all the $v$ variants, respectively. If $v=1$, no other compartments exist. If $v>1$, $\compart{seg}$ identifies plants infected by all the $v$ defective and complementing variants, without the \wt{}. In addition, there are $2^v-1$ other compartments containing $wt$ plus a combination of some (not all) of the defective variants. We name them according to which of the latter they are infected by. For instance, $\compart{1}$ contains \wt{} plus variant $1$, and $\compart{3,5}$ contains $wt$ plus variants $3$ and $5$. If defective variants are not present, $\compart{wt}$ behaves like a standard Susceptible--Infected--Susceptible (SIS) model, with probability of transmission upon contact equal to $\lambda$.

We implement enhanced transmissibility of defective variants by assuming they spread with a probability $\rho\lambda$. $\rho=1$ thus means that \wt{} and segments are epidemiologically equivalent, while any value larger than $1$ causes the defective variants to transmit more easily than the \wt{}. For a graphic representation of the spreading routes and differential transmissibility see Fig.~\ref{fig:illustration}D. In the general case, an agent in a given compartment may transmit some (or all) of the viral species it hosts to the one it is in contact with, with a probability depending on the initial compartments of the two agents, and on the final compartment. For instance, a host in compartment $\compart{1}$, upon contact with a susceptible one, may transmit both \wt{} and $1$, turning the susceptible into a $\compart{1}$. It may instead transmit \wt{} alone, turning the susceptible into a $\compart{wt}$. The defective variant, however, cannot be transmitted alone, as it requires \wt{}, as previously stated.
A schematic representation of the compartmental model for a bipartite virus ($v=2$) is depicted in Fig.~\ref{fig:illustration}C.

Our assumption of constant host population (of size $N$) holds for strictly constant size, as well as populations that are at equilibrium, i.e., the number of births equals the number of deaths, or at least any growth pattern occurs at time scales much larger than the spread of the virus. This assumption is connected to the spreading model, as the recovery process of the Susceptible-Infectious-Susceptible model can be regarded in two ways. It can be seen as proper recovery, with the host clearing the virus but acquiring no immunity to reinfection. It can also be interpreted as the virus killing the host, and a new (susceptible) host filling its ecological space.

In the absence of specific evidence~\cite{Sanjuan2017}, we make the simplest assumption for transmission: the different types of viral particles are transmitted independently, so that the probability of concurrent transmission of two variants (and \wt{}) is simply the product of the probabilities of the single events. We also assume that co-infection by \wt{} and variants does not alter the infectious period, allowing us to model recovery at a rate $\mu$ for all infected hosts.
In \textit{Methods} and~\nameref{S1_Text} we expand our analysis to account for nonindependent transmission, heterogeneous recovery rates, and the case when different variants compete for a limited carrying capacity within the host, due, for instance, to a limited number of viral particles a host cell can make per unit time. We show that all these additional features do not impact the qualitative behavior of our model, in agreement with what was previously found in~\cite{Iranzo2012}.

When the virus is introduced into a susceptible population, it can either die out quickly and leave the system disease-free (disease-free state, \dfs{}), or reach endemicity. There are four possible endemic states, depending of which variants circulate. We equivalently use the term {\itshape equilibria}, as they are the stable equilibrium points of the spreading dynamics. The first one is {\bfseries wt}, in which only the \wt{} is prevalent, and the defective variants have died out. This case maps into an effective SIS model for the compartment $\compart{wt}$. In the second one, {\bfseries hj}, any defective segment can circulate alongside the \wt{} because, roughly speaking, the transmissibility of the latter is so high that any defective variant can hijack it, with no need to complement the genome with other variants. In this case, we will likely see the circulation of a number of variants lower than $v$, as segments can go extinct without hampering the circulation of the remaining ones. The third endemic state, {\bfseries seg}, witnesses the presence of all the $v$ segmented variants without \wt{}, and in this case complementation is essential. This state is an SIS model for the compartment $\compart{seg}$. Finally, the state {\bfseries all} exhibits circulation of the \wt{} plus all the variants $v$.
Borrowing some terminology from physics, we can then define different epidemic phases. Phases are regions of the space of model parameters. Inside each phase, the macroscopic behavior of viral spread is qualitatively the same. Specifically, we can define a phase in terms of which endemic states it allows. The parameter surfaces separating different phases are called phase transitions, the most important in epidemiology being the epidemic threshold. Below the epidemic threshold, only the \dfs{} exists. Above it, the pathogen can circulate. There, we identify five other phases: \wt-phase allows only {\bfseries wt}; {\itshape contingent}-phase allows {\bfseries wt}, {\bfseries hj} and {\bfseries all}; {\itshape mix}-phase allows {\bfseries wt}, {\bfseries seg} and {\bfseries all}; {\itshape seg}-phase allows only {\bfseries seg}; finally {\itshape all}-phase admits all the possible endemic states.

Table~\ref{tab:fasi-equilibri} provides a schematic representation of the relationship between phases and endemic states.

\begin{table}[!ht]
	\centering
	\caption{
		{{\bfseries Connection between phases and endemic states}. Connection between the possible phases of the epidemic and the endemic states they allow. A tick mark connecting a phase and an endemic state means that the latter is a stable equilibrium in that phase, and can occur.}}
	\begin{tabular}{|r+c|c|c|c|}
		\hline
		\multirow{2}{*}{{\bf Phases}}  & \multicolumn{4}{c|}{\bf Equilibria (endemic states)}\\
		& {\bfseries wt} & {\bfseries hj} & {\bfseries seg} & {\bfseries all} \\ \thickhline
		{\itshape wt} & $\checkmark$ &  &  &  \\ \hline
		{\itshape seg} &  &  & $\checkmark$ &  \\ \hline
		{\itshape contingent} & $\checkmark$ & $\checkmark$ &  & $\checkmark$ \\ \hline
		{\itshape mix} & $\checkmark$ &  & $\checkmark$ & $\checkmark$ \\ \hline
		{\itshape all} & $\checkmark$ & $\checkmark$ & $\checkmark$ & $\checkmark$ \\ \hline
	\end{tabular}
	\begin{flushleft}
	\end{flushleft}
	\label{tab:fasi-equilibri}
\end{table}

In the following, we analytically derive the critical surfaces that separate the different phases in the space of the parameters. This means that, given specific values of the parameters, the possible outcome of the spread can be predicted, thus characterizing the conditions leading to the persistence of multipartitism, and its nature. Then, using numerical simulations, we study the equilibrium prevalences of the endemic states, and their probability of occurring, for a representative set of parameter values.

Firstly, however, we need to set up the theoretical modeling framework in terms of reaction-diffusion equations. For a generic $v$, we order the compartments by increasing number of viral species they contain, starting from $\compart{wt}$, and ending with $\compart{seg}$, $\compart{all}$. For instance, for $v=3$, this would be $\compart{wt}$, $\compart{1}$, $\compart{2}$, $\compart{3}$, $\compart{12}$, $\compart{13}$, $\compart{23}$, $\compart{seg}$, $\compart{all}$. Within the framework of heterogeneous mean field~\cite{PastorSatorras2001,Newman2002,PastorSatorras2015}, we divide hosts in classes according to their contact potential ({\itshape degree} in the language of networks), so that if two hosts have degree $k,h$, respectively, their contact rate will be the product $kh$ (in the absence of degree-degree correlations). We assume hosts with the same degree are equivalent, and consider the prevalence per degree class. To this end, we define the variable $x_\nu^k$ as the prevalence of compartment with index $\nu$ and degree class $k$, i.e., the fraction of the host population which has that degree, and finds itself in that compartment. In terms of $x_\nu^k$, the equations describing the evolution of the disease are
\begin{align}
	\dot{x}_\nu^k = -\mu x_\nu^k + &\frac{k}{\pan{k}} \sum_\beta
	\left[ \Gamma_{\nu\beta} \left( 1-\sum_\sigma x_\sigma^k \right) \right. \nonumber
	\\
	& \left. + \sum_\sigma \Lambda_{\nu\beta\sigma} x_\sigma^k \right]
	\left( \sum_h h p_\gamma(h) x_\beta^h \right)\,,
	\label{eq:general}
\end{align}
where $p_\gamma(k)$ is the probability of a host having degree equal to $k$. We consider the {\itshape homogeneous} case, where all hosts have the same degree, so that $p_{\gamma}(k) = \delta_{k,1}$ (with no loss of generality we set it to $1$), and a highly {\itshape heterogeneous} case, where $p_\gamma(k) = C_\gamma k^{-\gamma}$ is a power-law with exponent $\gamma$, and normalization constant $C_\gamma$.
We denote $\pan{k^m}$ as the $m$-th moment of the degree distribution, computed as $\pan{k^m} = \sum_k p_\gamma(k) k^m$, as usual. The term $\pan{k}$ appearing in Eq.~\ref{eq:general} is then the expected degree. The Greek indices $\beta, \nu, \sigma$, run on all the infectious compartments defined before. The susceptible compartment is not included, as the number of susceptible hosts is completely determined by the other compartments, thanks to the assumption of constant population size.
$\Gamma_{\nu\beta}$ is the rate of the transition $\compart{\beta}\compart{S}\rightarrow \compart{\beta}\compart{\nu}$, i.e., a transition affecting the prevalence of compartment $\compart{\nu}$ through a contact between a host in compartment $\compart{\beta}$ and a Susceptible. $\Lambda_{\nu\beta\sigma}$ encodes transmission rates among infected individuals, and specifically a transmission from $\compart{\beta}$ to $\compart{\sigma}$, that leads to the change of the prevalence of $\compart{\nu}$. The entries of $\Gamma_{\nu\beta},\Lambda_{\nu\beta\sigma}$ are functions of $\lambda$, $\rho$ and $v$.
Equation~(\ref{eq:general}) thus links the change in the number of hosts with a given degree, and in a compartment ($\dot{x}_\nu^k$), to one reaction and two diffusion processes. The first term, $\mu x_\nu^k$, represents the decrease due to hosts recovering back to the susceptible state. The second one, with coupling constant $\Gamma$, contains the probability of a host, with degree $k$, being susceptible ($1-\sum_\sigma x_\sigma^k$), and being infected by a host in compartment $\beta$, and degree $h$. The last term, with coupling constant $\Lambda$,  has the same structure, but the target compartment is a generic infectious compartment $\sigma$, instead of the susceptible one. Both infection terms contain the term $k h p_\gamma (h)/\pan{k}$, which is the probability a host of degree $k$ establishes a contact with a host of degree $h$, given a network with no degree-degree correlations~\cite{Barrat2008}.

The analytical approach to computing the critical surfaces consists in studying the linear stability of the different equilibria of Eq.~(\ref{eq:general}). Instead, in order to compute the prevalence values and occurrence probability of these equilibria, we have to resort to stochastic spreading simulations. The extensive calculations are reported in \textit{Methods} and~\nameref{S1_Text}, as well as the explanation of the numerical simulations.

\subsection*{Critical behavior}

\begin{figure}[htbp]
	\includegraphics[width=.65\textwidth]{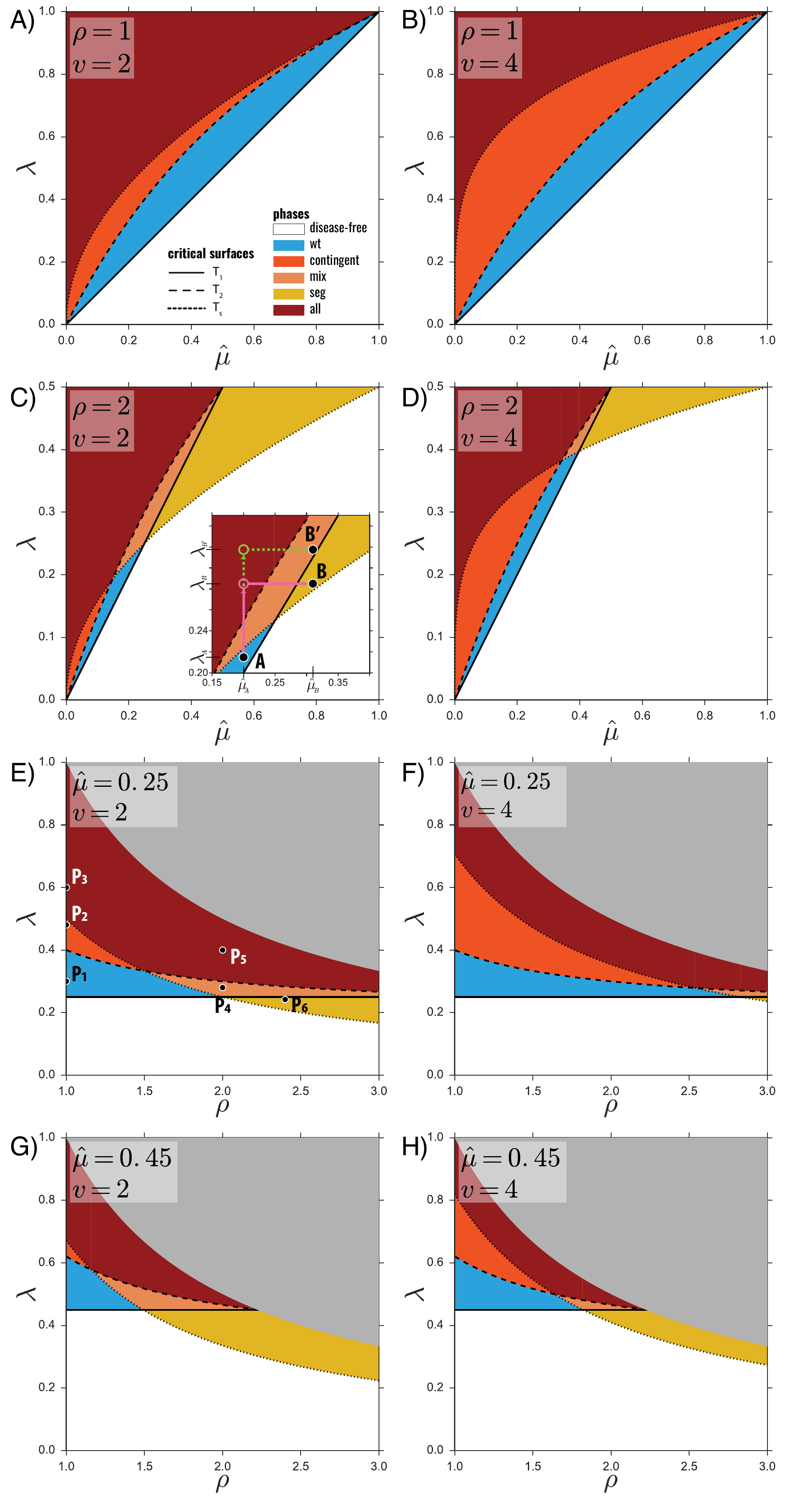} 
	\caption{Parameter exploration and endemic phases. Of the four parameters that influence the critical points ($\hat{\mu},\lambda,\rho,v$), two are in turn kept fixed and the remaining two explored in a two-dimensional plot highlighting the different phases. The value of the fixed parameters is reported on the top left of each plot. The $y$-axis of each plot is always the transmissibility of \wt{} ($\lambda$). The critical surfaces are $T_1,T_2,T_s$ (solid, dashed, dotted lines), and the phases are colored as in the legend. The gray areas indicate forbidden parameter values (probabilities higher than $1$). For a numerical validation of $E)$ see~\nameref{S1_Text}. The inset in $C)$ is a magnification of a subregion of the $C)$ plot. In $E)$ the points displayed have the following values: $P_1=(1,0.3)$, $P_2=(1,0.48)$, $P_3=(1,0.6)$, $P_4=(2,0.28)$, $P_5=(2,0.4)$, $P_6=(2.4,0.248)$.}
	\label{fig:panel}
\end{figure}

The five phases are completely determined by three surfaces with tractable analytical expressions. They are $T_1$, above which the \wt{} can spread on its own (epidemic threshold for the compartment $\compart{wt}$ while alone); $T_2$, above which segments circulate by hijacking the \wt{}, and $T_s$, which is the epidemic threshold for the compartment $\compart{seg}$ circulating alone. The expressions we find are
\begin{align}
	T_1 & = \left\{\lambda = \hat{\mu} \right\} ; \label{eq:T1_dd} \\
	T_2 & = \left\{ \lambda = \frac{1+\rho}{\rho} \frac{\hat{\mu}}{1+\hat{\mu}} \right\}; \label{eq:T2_dd} \\
	T_s & = \left\{ \lambda = \frac{\hat{\mu}^{1/v}}{\rho} \right\}. \label{eq:Ts_dd}
\end{align}
$\hat{\mu}$ is an effective recovery-rate embodying both the actual recovery rate, and the topology of the contacts: $\hat{\mu} = \mu \pan{k} / \pan{k^2}$. This entails an important scaling: recovery rate and topology never impact the critical points on their own, but always jointly as $\hat{\mu}$. This fact was well-known in the case of the epidemic threshold, Eq.~(\ref{eq:T1_dd})~\cite{PastorSatorras2001}. Here, we rigorously prove that it extends also to all other critical points. Given that homogeneous contact networks have $\pan{k}\sim \pan{k^2}$, the heterogeneous (power-law-like) network recovers the homogeneous case when $\gamma\rightarrow\infty$.
Hence, the smaller the exponent $\gamma$, the more heterogeneous the contact network is, i.e., hosts with a large number of connections become more likely. These hosts can reach a significant part of the population, and when infectious, they act as superspreades. They are responsible for causing $\hat{\mu}$ to go to zero ($\hat{\mu}\rightarrow 0$) in the limit of large population size ($N\rightarrow\infty$), when the exponent of the degree distribution respects $2<\gamma<3$.
This implies not only that $T_1$ goes to zero, as it is well-known~\cite{PastorSatorras2001}, but that $T_2$ and $T_s$ do it as well. However, while $T_2/T_1$ remains finite as $\hat{\mu}$ goes to $0$ ($T_1$ and $T_2$ go to zero at the same speed), we find that $T_s/T_1\rightarrow\infty$. This entails that $T_s$ goes to zero more slowly, and increasingly so for higher $v$.

The study of Eqs.~(\ref{eq:T1_dd})--(\ref{eq:Ts_dd}) reveals four regimes. For low or zero differential transmissibility ($\rho<\hat{\mu}^{-(v-1)/v} - (1-\hat{\mu}^{1/v})$), as $\lambda$ increases, one crosses the {\itshape wt}-phase, then the $contingent$-phase and finally the $all$-phase (see Fig.~\ref{fig:panel}A). For intermediate values of differential transmissibility ($\hat{\mu}^{-(v-1)/v} - (1-\hat{\mu}^{1/v}) < \rho < \hat{\mu}^{-(v-1)/v}$), the $mix$-phase substitutes the $contingent$-phase. This can be seen in Figs.~\ref{fig:panel}B and~\ref{fig:panel}C. Then, when $\hat{\mu}^{-(v-1)/v} < \rho < \hat{\mu}^{-1}$, increasing $\lambda$ causes the system to be in the $seg$-phase, followed by the $mix$-phase and later by the $all$-phase (see Figs.~\ref{fig:panel}B, \ref{fig:panel}C and~\ref{fig:panel}D). Finally, for very high differential transmissibility $\rho>\hat{\mu}^{-1}$, the \wt{} no longer spreads and the only possible phase is the $seg$-phase (see Figs.~\ref{fig:panel}B and~\ref{fig:panel}C).

\subsection*{Endemic prevalences}

For any possible value of the parameters, Eqs.~(\ref{eq:T1_dd})--(\ref{eq:Ts_dd}) tell us which endemic states are possible, i.e., which prevalences are higher than zero. They provide, however, no information about the values of such prevalences, which are, in principle, the solutions of the algebraic system obtained by setting $\dot{x}_\nu = 0$ in Eq.~(\ref{eq:general}). A closed-form solution of this system does not exist for heterogeneous networks. In the homogeneous case, while a complete analytical derivation of the endemic states is not possible, we can obtain two important results. Firstly, we notice that the total prevalence of the \wt{}, i.e., the fraction of hosts infected by it ($z=\sum_{\nu\not= \compart{seg}}x_\nu$), obeys an SIS dynamics (see~\nameref{S1_Text}) with transmissibility $\lambda$, and can thus be computed as $z_{wt}=1-\mu/\lambda$. Secondly, when the whole set of segments circulates without \wt{} (as compartment $\compart{seg}$), again the virus spreads as an SIS, this time with transmissibility $\left(\rho\lambda\right)^v$, and its endemic value can be predicted in the same fashion: $z_{seg} = 1-\frac{\rho}{\left(\rho\lambda\right)^v}$. Interestingly, for high $\rho$, and a transmissibility $\lambda > \rho^{-v/(v-1)}$, it turns out that $z_{seg}>z_{wt}$: the prevalence of the multipartite form is higher than that of the \wt{}.

In order to fully characterize the endemic states, we resort now to stochastic spreading simulations (see \textit{Methods}), focusing on the bipartite case ($v=2$). A higher number of variants ($v>2$) would not change the qualitatively picture; it would simply increase the possible values for the prevalence of {\bfseries hj} by increasing the number of possible segments that survive through hijacking. We choose six points in the parameter space that lie in different phases (see Fig.~\ref{fig:panel}E), and for those values we carry out the simulations.


\begin{figure}[htbp]
	\includegraphics[width=\linewidth]{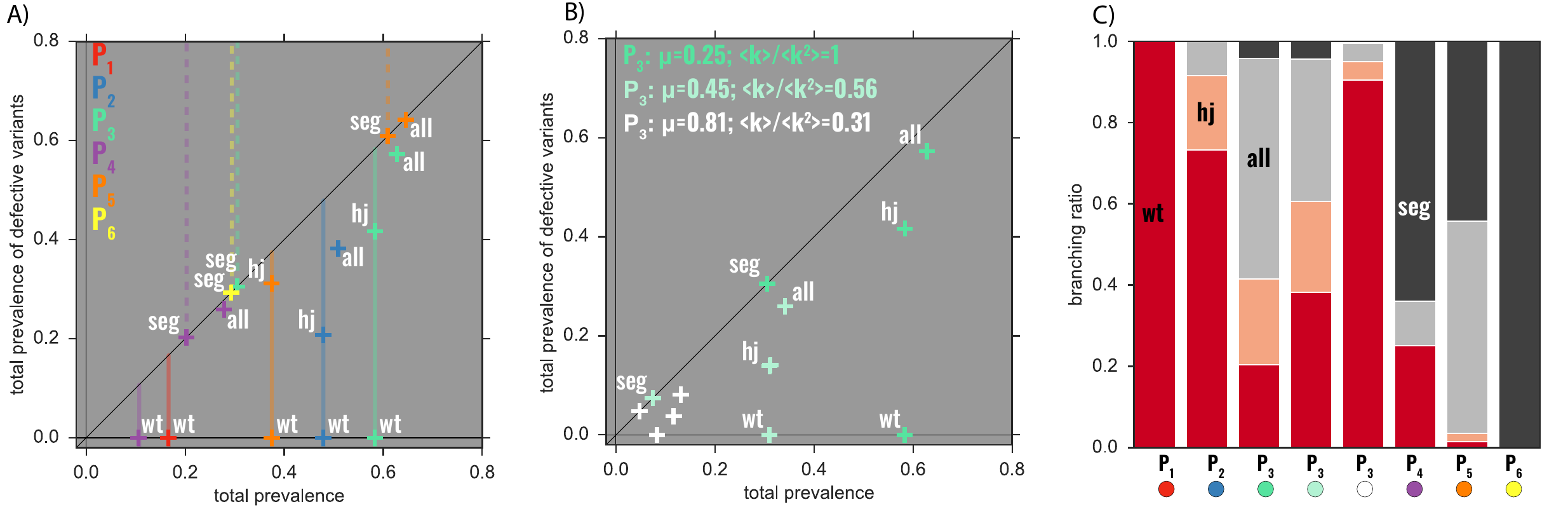} 
	\caption{Endemic states. Plots $A)$ and $B)$ show the results of the simulations for the endemic states of configurations corresponding to the points in Fig.~\ref{fig:panel}E, for a bipartite virus ($v=2$). In both $A)$ and $B)$, {\itshape x}-axis is the total prevalence of the disease, i.e., the fraction of hosts infected by any configuration of the virus, at equilibrium, and the {\itshape y}-axis is the prevalence of the defective variants, i.e., the fraction of hosts infected by at least one defective segment. The points are numerically recovered endemic states, their type being indicated by the labels.  In $A)$ the underlying contact network is homogeneous, so that $\hat{\mu}=\mu=0.25$. The solid vertical lines mark the analytical prediction of the total prevalence when \wt{} is present either alone or together with just one variant. The dashed vertical lines mark the analytical prediction of the prevalence when the segments circulate without \wt{}. The crosses mark the prevalence values of the equilibrium points averaged over the runs not leading to extinction, among the $5000$ executed per point. $B)$ focuses on $P_3$ (in Fig.~\ref{fig:panel}E): the fixed value $\hat{\mu}=0.25$ is obtained either as $\mu = \hat{\mu}$ (homogeneous network, as in (B)), or with two heterogeneous networks with exponent $\gamma=3.5,\mu=0.45$ and $\gamma=3.2,\mu\approx0.81$ (and thus $\pan{k}/\pan{k^2}\approx 0.56$ and $\pan{k}/\pan{k^2}\approx 0.31$, respectively).  For each of the points and the equilibria examined, $C)$ reports the branching ratio, defined as the probability of reaching that particular equilibrium. They are computed by starting all the simulations with all susceptible but one in $\compart{all}$ (infected by \wt{} and all the variants), and counting the fraction of the runs that reach that equilibrium, among the ones that do not go to extinction.}
	\label{fig:equilibri}
\end{figure}

We firstly focus on homogeneous host population structures. The results are shown in Fig.~\ref{fig:equilibri}A. We characterize the endemic states in terms of their type (see Tab.~\ref{tab:fasi-equilibri}), and plot their total prevalence, and the prevalence of the defective variants.
In the points lying on the $x$-axis (labeled by {\bfseries wt}) the defective variants have gone extinct, and the \wt{} behaves like an SIS (states {\bfseries wt}). The points lying on the diagonal have witnessed the extinction of the \wt{}, and the defective variants are circulating together in the $\compart{seg}$ compartment (states {\bfseries seg}). Their values match the theoretical prediction (dashed vertical lines). The solid vertical lines in Fig.~\ref{fig:equilibri}A are the theoretical predictions of \wt{} prevalence. They match all equilibria of type both {\bfseries wt} and {\bfseries hj}, as in those cases the total prevalence coincides with the prevalence of the \wt{}. The states {\bfseries all}, whose total prevalence cannot be predicted analytically, have the highest prevalence. This picture further confirms the relationship between the theoretically predicted phases and the allowed endemic states (Fig.~\ref{fig:panel}D).


Previously we have stated that the critical surfaces are not sensitive to recovery rate and topology separately, but only to the parameter $\hat{\mu}$ encoding both at the same time. Specifically, two populations with different recovery rate and contact heterogeneity, but with the same $\hat{\mu}$, are indistinguishable from the point of view of their critical behavior. The endemic prevalences, however, break this symmetry, as one can see from Fig.~\ref{fig:equilibri}B, where we focus on $P_3$ (Fig.~\ref{fig:panel}D) and get to $\hat{\mu}=0.25$ both with one homogeneous (as in Fig.~\ref{fig:equilibri}B), and  two heterogeneous population structures (with exponents $\gamma=3.5$ and $\gamma=3.2$). All the three configurations show all the equilibria, as expected by the critical behavior, but in the heterogeneous case the prevalence is consistently lower for each equilibrium.

\subsection*{Likelihood of different endemic states}

Up to now, we have identified the phases (allowed endemic states) and computed the prevalence of such states. We now focus on the probability of occurrence of each state. For each of the usual points in Fig.~\ref{fig:panel}E, we show the probability of reaching each equilibrium in Fig.~\ref{fig:equilibri}C. This is achieved by counting the number of stochastic realizations that, starting from similar initial conditions, lead to that specific equilibrium ({\itshape branching ratio} of that equilibrium). Clearly, points $P_1$ and $P_6$ have only one endemic state, which then has a probability equal to one of being reached. For the other points, which have more than one possible endemic scenario, these probabilities are more informative, as they tell us the chances of the different viral forms taking over the population. We remark, however, that while both the critical behavior and the prevalence of the endemic states are inherent properties of the system that do not depend on the specific initial conditions chosen, this is not true for the probabilities of occurrence, which are clearly influenced by the initial infection status of the population. Given that, however, we computed them by seeding only one host in the $\compart{all}$ compartment to a susceptible population, we can say that our predictions are---at least qualitatively---reliable in an invasion scenario, in which the viral form is introduced by just one (or few) individuals.


\section*{Discussion}

\subsection*{Rise of multipartitism}

Using the analytical characterization of the endemic phases and the numerical study of the equilibria, we now can investigate under which conditions the interplay between spreading dynamics and topology of contacts leads to the rise and persistence of multipartitism. We can also determine the nature of such emergence, in terms of a commensal relationship with the \wt{}, or a true competitive advantage at the ecological level. For the sake of simplicity, we start by considering no differential transmissibility ($\rho=1$): \wt{} and segments have the same transmission probability. The relevant figures are Figs.~\ref{fig:panel}A and~\ref{fig:panel}B, points $P_1$, $P_2$ and $P_3$ in Fig.~\ref{fig:panel}E and Fig.~\ref{fig:equilibri}. In this scenario, three phases are possible, and one crosses them all by increasing the transmissibility $\lambda$. The first one ($\lambda$ just above $T_1$) is the \wt-phase, in which only the \wt{} can circulate, and whenever a defective segment is produced, it quickly goes extinct. By increasing $\lambda$, we then encounter the {\itshape contingent}-phase. This phase predates the appearance of true multipartitism, as defective segments can hijack the \wt{} to circulate. These segments cannot persist on their own, but the highly prevalent \wt{} allows them to complete the replication cycle. At this stage, any defective segment is a commensal of \wt{}, as the persistence of the former depends on the latter, while \wt's fitness remains unchanged. The emergence of multipartitism in this context is a contingent process: segments circulate simply because they are allowed to, causing no change to the overall fitness of the virus. Furthermore, there is no selective pressure towards complementation, as a combination of segments reconstructing the full genome without the presence of the \wt{} (compartment $\compart{seg}$) would not be able to persist. This is confirmed by the functional form of $T_2$ in Eq.~(\ref{eq:T2_dd}), which features no dependence on the number of complementing variants $v$: the survival of each mutant is independent of the presence of others, as effective replication and diffusion is \wt{}-mediated. In other words, complementation would not make the variants fitter to the environment.

A further increase in $\lambda$ takes us to the {\itshape all}-phase. Here, in addition to the commensal relationship between \wt{} and segments, complementing variants are able to circulate on their own, without \wt{}: the equilibrium {\bfseries seg} emerges. Selection then imposes a bias on those segments that together reconstruct the genome, as they represent a new effective spreading configuration, and increase the overall viral fitness. They are thus advantaged with respect to purely commensal segments. This fitness-enhancing effect is quite straightforward: let us suppose that, due to viral or host population bottlenecks, or another stochastic event, \wt{} prevalence goes down drastically, to the point where it is cleared from the system. In the {\itshape contingent}-phase this would lead to complete viral extinction, as all the segments would die out, too, as their persistence is linked to \wt's. In the {\itshape all}-phase, on the other hand, the virus is more resilient, as it can still circulate thanks to complementation. This time  selective pressure toward complementation is well visible in the expression of $T_s$ in Eq.~(\ref{eq:Ts_dd}), which depends exponentially on the number of variants $v$. This fact is in qualitative agreement with results in~\cite{Iranzo2012}, where it was shown that the larger the number of segments, the harder to reach the persistence of the segmented variants within a host. Even if the multipartite genome does not enjoy any microscopic advantage, it can rise to fixation if the monopartite virus undergoes stochastic extinction. Though fluctuations would also affect the multipartite form, and stochastic extinction of the monopartite form is not very likely, similar scenarios are relevant in virus evolution \cite{grande2005suppression,iranzo2008stochastic} and cannot be discarded {\it a priori}.

\subsection*{The {\itshape contingent}-phase: a stepping stone towards multipartitism}

The analysis of the prevalences (Fig~\ref{fig:equilibri}A) confirms the evolutionary drivers behind the different phases, and adds information regarding crossed effects between viral types. Moreover, it allows us to uncover an evolutionary potential for multipartitism even in the absence of an explicit microscopic advantage. Let us focus on the {\itshape contingent}-phase (point $P_2$ in Fig~\ref{fig:equilibri}A), and the {\bfseries all} state. The total prevalence of the latter state is higher than {\bfseries wt}'s and  {\bfseries hj}'s in the same phase, implying that some hosts are infected by complementing segments without \wt{} (compartment $\compart{seg}$), that is by a {\it bona fide} multipartite virus. Given that the multipartite form is not endemic in this phase---the {\itshape contingent}-phase relies on the \wt{} for viral persistence---, this excess prevalence of the {\bfseries all} state is a by-product of overall viral prevalence, rather than its driver: an extinction of the \wt{} would quickly drive segmented variants to extinction. It is, however, an important one, as while it may not increase fitness in that specific environment, it permits the independent replication of the set of complementing variants; these are then able to invade other environments in which the \wt{} could not persist, as we will see in the following.

\subsection*{Contact heterogeneity impairs multipartitism}

Let us examine the effect of a heterogeneous contact network on the phases and equilibria above. As we have explained, in the phase space, topology is encoded in the parameter $\hat{\mu}$. A low epidemic threshold is a well known feature of heterogeneous networks~\cite{PastorSatorras2001,Newman2002,Barrat2008}. Specifically, power-law networks with $\gamma<3$ exhibit a vanishing threshold as they grow larger, as the emergence of highly connected hubs ensures the persistence of the disease at any value of transmissibility, that is $\pan{k}/\pan{k^2}\rightarrow 0$ (and as a result, $\hat{\mu}\rightarrow 0$) as the number of potential hosts grows, $N\rightarrow\infty$. In our case this translates into $T_1$, which is the epidemic threshold, going to zero for $\hat{\mu}\rightarrow 0$. Also $T_2,T_s\rightarrow 0$. Further information is obtained when comparing their limit behaviors. As $\hat{\mu}$ becomes smaller, $T_2/T_1$ increases but remains finite, while $T_s/T_1 \rightarrow\infty$. This implies that, the higher the heterogeneity of the network is, the more difficult it becomes for multipartitism to persist. Specifically, reaching the {\itshape all}-phase from the {\itshape wt}-phase would require an infinite relative increase (in the limit $\hat{\mu}\rightarrow 0$) in transmissibility. Even when heterogeneity is not severe enough so as to cause the threshold to vanish, i.e., when $\gamma>3$, heterogeneity makes it harder to sustain multipartitism, as both $T_2/T_1$ and $T_s/T_1$ are decreasing functions of $\hat{\mu}$.

Heterogeneity also modifies endemic prevalences and the branching ratios of equilibria. Let us examine point $P_3$ ({\itshape all}-phase in Fig.~\ref{fig:panel}E): when the network is homogeneous, the highest branching ratio corresponds to {\bfseries all}, and the equilibria containing segments together ({\bfseries hj}) happens $20\%$ of the time. When the network is heterogeneous, this fraction decreases, and {\bfseries wt} quickly overtakes {\bfseries all} in being the most probable outcome.

Summarizing, homogeneous contact patterns favor the emergence and persistence of multipartitism, while heterogeneous contacts hamper it. Qualitatively, this is the result of the complex interaction between the bottlenecks induced by between-host transmission and the presence of superspreaders, i.e., hosts that can potentially infect a large fraction of the population thanks to their high number of contacts. Combining this mechanism with a low MOI---and hence low $\lambda$---predicts an evolutionary radiation of multipartite viral forms linked to the rise and intensification of agricultural practices. In crops and cultivars, contacts among hosts are much more homogeneous (and often closer) than in wild settings, tremendously alleviating the requirements imposed by co-infection. Multipartite viruses adapted to the patchy distribution of wild hosts could have found it easy to propagate in regular, monospecific host populations which in all cases have closely related wild forms from which they departed through artificial selection~\cite{meyer:2013}.

\subsection*{Emergence of new multipartite phases through increased transmissibility of the segmented form}

Up to this point, we have assumed that all viral forms have the same transmissibility and, still, fixation of the multipartite form cannot be fully discarded. Any additional advantage, however minor, of multipartitism will contribute to its ecological success, as we now discuss. We now set out to study the impact of enhanced transmissibility of defective variants, encoded in the parameter $\rho$ being larger than one ($\rho>1$). As $\rho$ increases, the endemic state {\bfseries seg} becomes more prevalent and more likely with respect to {\bfseries wt} (see Figs.~\ref{fig:equilibri}C and~\ref{fig:equilibri}D). Specifically, the value of the transmissibility for which $z_{seg}>z_{wt}$ decreases as $\rho$ increases, facilitating the predominance of the multipartite form (in Fig.~\ref{fig:equilibri}A point $P_3$ has $z_{wt}<z_{seg}$, while $P_4$ and $P_5$ have $z_{seg}>z_{wt}$). Most importantly, $\rho>1$ causes two new phases to emerge (Figs.~\ref{fig:equilibri}B, \ref{fig:equilibri}C and ~\ref{fig:equilibri}D), and both facilitate the rise of multipartitism by eliminating {\bfseries hj} from their possible equilibria. One is the {\itshape mix}-phase, in which the virus circulates either as \wt{} or as a multipartite. The {\itshape mix}-phase also presents an {\bfseries all} endemic state that results from the interaction between the two former equilibria. Unlike in the {\itshape all}-phase, however, here the {\bfseries all} equilibrium no longer indicates commensal relation. The second emerging phase is the {\itshape seg}-phase, in which only the complemented multipartite virus is able to circulate, while the monopartite version quickly goes extinct (yellow, and point $P_6$ in Figs.~\ref{fig:panel} and~\ref{fig:equilibri}). This phase is of paramount importance because it lies in a parameter region where, without developing multipartitism, the virus would not be endemic. In addition, by prescribing the exclusive presence of the multipartite form, it allows to explain the phenomenology observed in nature, as the simultaneous presence of monopartite and defective-complementing forms of the same virus has been observed only {\it in vitro}~\cite{oneill:1982,geigenmuller:1991,kim:1997,garcia-arriaza:2004}. {\it In vivo}, viral species circulate as either pure monopartite (endemic state {\bfseries wt}), or pure multipartite ({\bfseries seg}). Furthermore, although {\it in vitro} several defective viral forms are generated and detected, and propagate along the \wt{} (which would correspond to an {\it in vitro} {\bfseries hj}), this equilibrium is rarely found in wild plants. There is, however, an association between fully-fledged viruses and defective viral forms formally equivalent to the {\bfseries hj} equilibrium: virus and viral satellites~\cite{kassanis:1962}. Often, in addition, satellites modify the aetiology of viral infections~\cite{rehman:2009}, such that the transmissibility and the recovery rate might be affected by its presence in no particular direction, a phenomenon we do not consider in our model. There are other classes of hyperparasites that depend on a functional virus for replication (e.g. virophages~\cite{scola:2008} or viroid-like satellites~\cite{flores:2011}) whose ecological dynamics could, with appropriate modifications, be described in the framework discussed here. Interestingly, it has been proposed that virus-satellite associations, a typically unrelated tandem from a phylogenetic viewpoint, might evolve towards full co-dependence, and therefore be a possible, alternative evolutionary pathway to multipartitism~\cite{Sanz2017}.

Though our knowledge of existing viral forms is still incomplete and likely biased~\cite{wren:2006,bernardo2017geometagenomics}, our results indicate that endemic states mixing monopartite and multipartite cognate forms ({\bfseries hj}, {\bfseries all}) need values of transmissibility difficult to sustain: endemicity could be achieved with lower values of transmissibility if the virus propagated only as a wild-type, while high values entail a cost that is usually compensated by decreasing infectivity~\cite{alizon:2009}. Albeit rare, however, these endemic states might act also as a stepping stone towards multipartitism even if they are only transiently present, as in the following example.

Consider a purely monopartite virus endemic in a plant population, in a specific environment, with parameters $\hat{\mu}_A$, $\lambda_A$ as in point $A$ in the inset of Fig.~\ref{fig:panel}C. $\hat{\mu}_A$ is a combination of the recovery rate of the disease (characteristic of the host-virus interaction), and the between-plant contact network, driven by plant distribution and vector movements. A second population, occupying an adjacent geographic area, may have a different parameter ($\hat{\mu}_B>\hat{\mu}_A$), due to a different contact topology. As the inset of Fig.~\ref{fig:panel}C shows, the virus is able to colonize the second population only through an evolutionary process that increases its transmissibility up to at least $\lambda_{B'}$, so that point $B'$ is above the epidemic threshold (green path in the figure). It is reasonable to assume that the larger the increase in transmissibility required, the less likely this process is, given that the required mutation(s) are less likely and possibly more costly to maintain. The emergence of multipartitism decreases the evolutionary distance between the two states, increasing adaptability (magenta path in the figure). Random mutations, in fact, need to increase transmissibility from $\lambda_A$ ({\itshape wt}-phase) to $\lambda_B<\lambda_{B'}$ ({\itshape all}-phase), where a complementing, multipartite version of the virus can emerge. Invasion of the second population is now possible, because the new viral forms effectively lowers the epidemic threshold in $\hat{\mu}_B$, thanks to the emergence of the {\itshape seg}-phase (point $B$). This simplistic example not only shows that multipartitism can emerge as a fitness-enhancing feature, but also that coexistence of monopartite and multipartite forms is a key stage in the evolutionary process, albeit possibly transient and short-lived. In addition to outlining the adaptive potential of multipartitism, this example elucidates the hampering effect of network heterogeneity. By increasing the distance between the {\itshape wt}-phase and the {\itshape all}-phase, network heterogeneity reduces the ratio $\lambda_{B'}/\lambda_{B}$, making multipartitism less advantageous. In conclusion, while making viral persistence overall easier, network heterogeneity curbs the potential of multipartitism as an effective adaptation strategy.


\section*{Conclusion}

Multipartitism represents an example of a complex and as-of-today puzzling viral strategy. We have developed a framework that, starting from few key biological features, models the interaction between monopartite and multipartite forms, driven by the spreading dynamics on a host population. Despite assuming that multipartitism emerged from complementation between defective viral forms generated by the \wt{} virus, as it has been observed {\it in vitro}, our results can be extended to other situations with relative ease. Most importantly, in addition, we have described how the structure of contacts among hosts drives the rise and persistence of the different viral forms. We have analytically characterized the parameter regions leading to viral persistence, in the form of \wt{} only, of \wt{} and defective segments, or segments only. We have also defined the different types of relationships between \wt{} and segments, and specifically the presence or absence of selective pressure towards complementation, i.e., to witnessing the circulation of defective variants that may cooperatively reconstruct the whole genome. We have corroborated these findings through stochastic numerical simulations aimed at computing the prevalence of the different endemic states, and their probability of occurrence. As a result, we have been able to identify under which ecological conditions would multipartitism be a successful adaptive strategy, in the presence or absence of microscopic advantages, to new external conditions and environments characterized by variations in the topology of contacts between hosts. Defective particles generated through replication errors would start circulating by hijacking the \wt{}. Subsequently, a complementing set of variants might form. Once that situation is achieved, even a small advantage in transmissibility ($\rho>1$) would give an advantage to the multipartite form, which could anyway replace the monopartite form if chance causes the stochastic extinction of the latter. This sequence of events represents a plausible, parsimonious evolutionary pathway to the rise and persistence of multipartite viruses, and clarifies in which manner multipartitism might be an effective adaptive strategy at the ecological level.

We have also uncovered that while heterogeneous contact patterns among hosts favor viral persistence in general, they give a higher advantage to monopartite forms, by limiting the evolutionary and adaptative potential of multipartitism. Our model clearly lacks specific biological features that characterize different viruses, but that is a strength rather than a weakness, as it can be applied to a wide variety of settings with appropriate minimal modifications. We nonetheless explore additional realistic features in \nameref{S1_Text}, as nonhomogeneous recovery rates and nonindependent viral transmission.

Finally, it is worth discussing the effect of the interaction between the microscopic advantage and stochastic effects on multipartite fixation.
The effect of the microscopic advantage, as quantified by our parameter $\rho$, becomes apparent in our current results, in terms of a much larger region of parameter space compatible with the fixation of the multipartite form (compare, for instance Fig.~\ref{fig:panel}B and Fig.~\ref{fig:panel}C).
Assuming a microscopic advantage therefore leads to a competitive advantage of multipartitism. A quantitative estimate of this competitive advantage, however, would require accounting for stochastic effects, an endeavor that goes beyond the current approach.

Despite not being able to formulate quantitative predictions, we are convinced that our framework provides an interesting qualitative picture of coexistence or substitution of different genomic architectures in a wide range of ecological environments. In this sense, we have uncovered evidence that the topology of contacts along which viruses spread may contribute to explaining why multipartite viruses preferentially infect plants. Our results lead us to conjecture that multipartite diversity and prevalence should have significantly increased together with the expansion of agriculture.

\section*{Materials and methods}

Our goal is to derive the critical surfaces of Eqs.~(\ref{eq:T1_dd})--(\ref{eq:Ts_dd}) from the equation driving the dynamics of the system, namely Eq.~(\ref{eq:general}). We do that by starting from a simpler scenario, and incrementally adding features, up to the full model. Specifically, the first step consists in solving the model with no differential degradation ($\rho=1$), no contact heterogeneity, and with only one segmented variant ($v=1$). In the second step we generalize the result to a generic $v$, and in the third one we allow for heterogeneous contacts. In the last step we add differential degradation. A numerical validation of the critical surfaces is carried out in~\nameref{S1_Text}.


$\compart{wt}$ and $\compart{all}$ are the only infectious compartments, with prevalence $x_1$ and $x_2$, respectively. With neither differential degradation nor contact heterogeneity, Eq.~(\ref{eq:general}) reduces to
%
%
\begin{equation}
	\left\{
	\begin{array}{rl}
		\dot{x}_1 = &\lambda (1-x_1-x_2)x_1 + \lambda(1-\lambda)(1-x_1-x_2) x_2 \nonumber
		\\
		&-\lambda x_1 x_2  - \mu x_1\,,
		\\
		\dot{x}_2 = &\lambda^2 (1-x_1-x_2) x_2 +\lambda x_1 x_2  - \mu x_2\,.
	\end{array}
	\right.
\end{equation}

%
By summing these equations, we find that the equation for the total prevalence ($z\udef x_1+x_2$) is $\dot{z} = \lambda (1-z)z -\mu z$. This also follows from noticing that for $v=1$ the total prevalence is also the total \wt{} prevalence (see~\nameref{S1_Text}). This means that the total prevalence behaves as a standard SIS, for which we know the epidemic threshold $T_1=\{\lambda=\mu\}$, and the equilibrium above it. In addition, we know that just above $T_1$ we are in the $wt$-phase. Hence, we have $\left(z_{wt}=1-\mu/\lambda, x_{2,eq}=0\right)$. Studying the stability of this equilibrium gives us $T_2$. Since the equation for $z$ decouples from $x_1$ and $x_2$, it is convenient to study the system in $(z,x_2)$. Studying the sign of the eigenvalues of the Jacobian matrix reduces to studying $\partial \dot{x}_2 / \partial x_2<0$ calculated in the equilibrium. This gives $T_2=\{\lambda=2\mu/(1+\mu)\}$. The details of the calculation are reported in the~\nameref{S1_Text}.


We now generalize the previous result to an arbitrary $v$, while still assuming that all hosts have the same contact rate, that we can set to one with no loss of generality. Eq.~(\ref{eq:general}) simplifies to
\begin{equation}
	\dot{x}_\nu  = \sum_{\beta\sigma} \Lambda_{\nu\beta\sigma} x_\beta x_\sigma + \sum_\beta \Gamma_{\nu\beta} x_\beta \left( 1-\sum_\sigma x_\sigma \right) - \mu x_\nu,
	\label{eq:general_homo}
\end{equation}
whose Jacobian matrix is
%
%
\begin{align}
	J_{\nu\beta} = \frac{\partial \dot{x}_\nu}{\partial x_\beta}
	= &\sum_\sigma \left( \Lambda_{\nu(\beta\sigma)} - \Gamma_{\nu\sigma} \right) x_\sigma \nonumber
	\\
	&+ \Gamma_{\nu\beta} \left( 1-\sum_\sigma x_\sigma \right) - \mu \delta_{\nu\beta}\,,
	\label{eq:general_homo_jacobian}
\end{align}
where $\Lambda_{\nu(\beta\sigma)} = \Lambda_{\nu\beta\sigma}+\Lambda_{\nu\sigma\beta}$. Firstly, we note that for $v>1$ the total prevalence, now defined as $z=\sum_{\nu} x_\nu$, no longer behaves like an SIS, due to the presence of the compartment $\compart{seg}$. Indeed, one can show that, when summing over $\nu$ in Eq.~(\ref{eq:general}), the terms with $\Lambda$ cancel out, as they pertain to interaction exclusively among infectious compartments, which by definition cannot change the total prevalence, and so all the contributions must cancel out. This is not the case however for the terms with $\Gamma$, so that the final equation is $\dot{z} = (1-z)\sum_\beta (\Gamma x)_\beta - \mu z$, which does not decouple from $x_\nu$. Interestingly, despite this breaking of the SIS symmetry, which was crucial to solve the $v=1$ model, we can still prove that the values of $T_1,T_2$ found for $v=1$ generalize to an arbitrary number of variants. We start from the first critical surface ($T_1$). We compute the Jacobian matrix of Eq.~(\ref{eq:general_homo_jacobian}) in the \dfs{}, i.e., $x_\beta = 0\; \forall\beta$. We get $J^{(\dfsm)} = \Gamma - \mu$. We now argue that $\Gamma$, and therefore the Jacobian matrix, is upper triangular, thanks to the specific ordering of the compartments that we introduced. $\Gamma_{\nu\beta}$ is the rate at which a susceptible becomes a $\compart{\nu}$, upon contact with a $\compart{\beta}$. For this to happen ($\Gamma_{\nu\beta}>0$), $\compart{\beta}$ must contain at least all the viral species $\compart{\nu}$ contains. Hence, either $\beta=\nu$ (diagonal term), or $\beta>\nu$. By the same reasoning, the diagonal terms are $\Gamma_{\beta\beta} = \lambda^{\phi(\beta)}$, where $\phi(\beta)$ is the number of viral species in $\compart{\beta}$, e.g. $\phi(\compart{wt})=1, \phi(\compart{all})=v+1$. From these considerations, the spectrum of $J^{(\dfsm)}$ is $\left\{ \lambda^{\phi(\beta)}-\mu; \, \forall\beta \right\}$. Keeping in mind that $\lambda<1$, we recover the first critical point: $T_1=\left\{\lambda=\mu\right\}$.

Just above $T_1$, $\compart{wt}$ is the only compartment with prevalence different from zero, hence it behaves like a standard SIS. Thanks to that we can compute Eq.~(\ref{eq:general_homo_jacobian}) in the $wt$-phase, and its spectrum. From that we find that the second critical surface is the same as for $v=1$. The details of the calculation are in~\nameref{S1_Text}.


We now build on the previous results, by adding heterogeneous contact rates. We work in the widely-used degree-block approximation~\cite{PastorSatorras2001,PastorSatorras2001b,Newman2002,Barrat2008,PastorSatorras2015}, assuming the contacts among agents are represented by an annealed network in which we assign each node a degree sampled from a power-law distribution with exponent $\gamma$: $p_\gamma(k) = C_\gamma k^{-\gamma}$, where $C_\gamma$ is the normalization factor. As customary, we assume $\gamma>2$, so that the average degree is defined in arbitrary large populations. In the framework of annealed networks it makes sense to interpret $k$ as a discrete number; one could also interpret it as a (continuous) coupling potential (either choice does not change the result found). We now directly compute the Jacobian of Eq.~(\ref{eq:general}), reported in Eq.~(S.16) of~\nameref{S1_Text}. The Jacobian is a matrix acting on a space which is the tensor product of the space of compartments, spanned by the Greek indices, and the space of degrees, spanned by the Latin indices. We can study its spectral properties on each space separately, using the previous results for the space of compartments. The full derivation is reported in the~\nameref{S1_Text}.


Differential degradation $\rho>1$ changes the matrices $\Gamma,\Lambda$, as reported in the~\nameref{S1_Text}. The derivation is then similar to the case with $\rho=1$.


Our model assumes that the transmission probability of one variant does not depend on the coinfecting variants. In reality, however, the number of viral particles a cell can produce in time is limited, and they are known to often spread in superinfection units. In \nameref{S1_Text} we investigate these aspects using a simple assumptions. We show that despite altering the specific values of the critical surfaces, they do not impact the qualitative behavior of the model.


Data in Fig.~\ref{fig:equilibri} are produced through stochastic spreading simulations. Starting with a population of $N=6000$, we infected the hosts with \wt{} and both segments ($\compart{all}$ for $v=2$), and let the virus spread. We used an adaptation of the Gillespie algorithm~\cite{Doob1942,Vestergaard2015}, to model both contacts among hosts, and contagion and recovery events. For each parameter configuration, we carried out $5000$~simulations and kept only those reaching an endemic state other than the disease-free state, in order to discard instances of stochastic extinction, and focus only on the metastable equilibria which represent the attractors of the equations. We then used those simulations to compute prevalences and occurrence probabilities.

\section*{Supporting information}

\paragraph*{S1 Text}
\label{S1_Text}
{\bf Detailed explanation of calculations and simulations.} We provide a thorough explanation of the analytical findings. We start from only one segmented variant ($v=1$) in Section~1, then generic $v$ (Section~2), and subsequently incorporate contact heterogeneity (Section~3). In Section~4 we include differential transmissibility of segmented variants. In Section~5 we analytically derive the total prevalence of the wild-type form. In Section~6, we provide a numerical validation of the critical surfaces in the phase space. In Section~7, we derive the critical surfaces when accounting for limited viral production by host cells. Finally, in Section~8 we consider simple corrections to our model accounting for nonhomogeneous recovery rates, and nonindependent transmission probabilities.

\section*{Acknowledgments}
We thank Michele Re Fiorentin for useful computational support and Carolyn M. Malmstr\"om and Israel Pag\'an for useful discussions. AA, EV and SG have been supported by the Generalitat de Catalunya project 2017-SGR-896, Spanish MINECO project FIS2015-71582-C2-1, and Universitat Rovira i Virgili project 2017PFR-URV-B2-41. SM acknowledges support from Spanish MINECO project FIS2017-89773-P. AA acknowledges financial support from the ICREA Academia and the James S.\ McDonnell Foundation.


%
%
%


\mbox{}

\mbox{}

\section*{Supporting information}

\setcounter{figure}{0}
\setcounter{section}{0}
\setcounter{equation}{0}
\renewcommand\thefigure{S.\arabic{figure}}
\renewcommand{\theequation}{S.\arabic{equation}}
\renewcommand{\thesection}{S.\arabic{section}}

\section{Model $v=1$}

Here we consider $v=1$, homogeneous contacts, and no differential degradation. Let $x_1,x_2$ be the prevalence of $\compart{wt},\compart{all}$, respectively. Equation~(\ref{eq:general}) reduces to
\begin{equation}
\begin{cases}
\
 \dot{x}_1 = -\mu x_1 + \lambda (1-x_1-x_2) x_1 + \lambda (1-\lambda) (1-x_1-x_2) x_2 - \lambda x_1 x_2; \\
 \dot{x}_2 = -\mu x_2 + \lambda^2  (1-x_1-x_2) x_2 + \lambda x_1 x_2.
\end{cases}
\end{equation}

As explained in the main text, the equation of the total prevalence $z=x_1+x_2$ decouples (see also Sect.~\ref{sec:totalwtprevalence}). It is thus convenient to consider the system in $(z,x_2)$:
\begin{equation}
\begin{cases}
 \dot{z} = \lambda (1-z)z -\mu z \\
 \dot{x}_2 = \lambda^2 (1-z) x_2 +\lambda (z-x_2) x_2  - \mu x_2.
\end{cases}
\label{eq:v1}
\end{equation}
As the equation for $z$ decouples from $x_2$, the Jacobian is lower triangular:
\begin{equation}
 J =
\begin{pmatrix}
  -\mu + \lambda\left(1-2z\right) & 0 \\
  \lambda (1-\lambda) x_2 & \lambda^2(1-z) + \lambda (z-2x_2) - \mu
\end{pmatrix}.
\end{equation}
The spectrum of $J$ is thus given by its diagonal elements. In order to get $T_1$, i.e., the epidemic threshold, we need to study the spectrum of $J$ computed in the disease-free state ($z=x_2=0$):
\begin{equation}
 J^{(\dfsm)} =
\begin{pmatrix}
  -\mu + \lambda & 0 \\
  0 & \lambda^2 - \mu
\end{pmatrix}.
\end{equation}
From this we see that if $\lambda>\mu$ the \dfsi\ is no longer stable. Hence $T_1=\{\lambda=\mu\}$. One could guess this without calculations from the equation in $z$, which tells us that the total prevalence behaves like an SIS. In order to find $T_2$ we now study the stability of the equilibrium where only $wt$ is circulating (hosts in $\compart{wt}$, but not in $\compart{all}$, are present). This is the equilibrium {\bfseries wt} defined in the main text, and it is a pure SIS model for the compartment $\compart{wt}$. The value of the prevalence is $(z=1-\mu/\lambda, x_2=0)$, as the SIS prescribes. The Jacobian in this equilibrium point is
\begin{equation}
 J^{(\mathbf{wt})} =
\begin{pmatrix}
  \mu-\lambda & 0 \\
  0 & \lambda (1+\mu)-2\mu
\end{pmatrix}.
\end{equation}
The first eigenvalue is always negative, as we are above $T_1$. The second one is negative iff $(1+\mu)\lambda < 2\mu$. As a result, we get that $T_2=\{\lambda=2\mu/(1+\mu)\}$.

\section{Generic number of variants $v$}
\label{sec:generic}

Assuming a generic number of variants, and homogeneous contacts, Eq.~(\ref{eq:general}), and its Jacobian, are
\begin{equation}
 \dot{x}_\nu  = \sum_{\beta\sigma} \Lambda_{\nu\beta\sigma} x_\beta x_\sigma + \sum_\beta \Gamma_{\nu\beta} x_\beta \left( 1-\sum_\sigma x_\sigma \right) - \mu x_\nu;
 \label{seq:general_homo}
\end{equation}
\begin{equation}
J_{\nu\beta}  = \frac{\partial \dot{x}_\nu}{\partial x_\beta} = \sum_\sigma \left[ \Lambda_{\nu(\beta\sigma)} - \Gamma_{\nu\sigma} \right] x_\sigma  +\Gamma_{\nu\beta} \left( 1-\sum_\sigma x_\sigma \right)     - \mu \delta_{\nu\beta},
\label{seq:general_homo_jacobian}
\end{equation}
with $\Lambda_{\nu(\beta\sigma)}=\Lambda_{\nu\beta\sigma}+\Lambda_{\nu\sigma\beta}$. They correspond to Eqs.~(5) and~(6) of the main text, respectively. 

As in the case $v=1$, we use the Jacobian, Eq.~(\ref{seq:general_homo_jacobian}), to study the stability of two equilibria. The first one is the \dfs{} ($x_\nu=0$), whose analysis gives $T_1$. The second one is {\bfseries wt}: $x_{1}=1-\mu/\lambda, \; x_\nu=0$ for $\nu>1$, and will give $T_2$. We remind that, given the ordering we use, the index $\nu=1$ refers to the compartment $\compart{wt}$, which is indeed the only one with non-zero prevalence in the {\itshape wt-phase} equilibrium. We study the stability of the former directly in the main text, so here we directly proceed to the latter.

The Jacobian computed in {\bfseries wt} is
\begin{equation}
 J^{(\mathbf{wt})}_{\nu\beta} = \left(1-\frac{\mu}{\lambda}\right)\left( \Lambda_{\nu(\beta 1)} - \Gamma_{\nu 1} \right) + \frac{\mu}{\lambda}\Gamma_{\nu\beta} -\mu\delta_{\nu\beta} .
\end{equation}
We can write it in matrix form by defining the following matrices: $\left(\Lambda\right)_{\nu\beta} = \Lambda_{\nu(\beta 1)}$, $\left(\tilde{\Gamma}\right)_{\nu\beta} = \Gamma_{\nu 1}$. The result is
\begin{equation}
 J^{(\mathbf{wt})} = \Gamma - \mu + \left(1-\frac{\mu}{\lambda}\right) \left(\Lambda - \Gamma - \tilde{\Gamma}\right).
 \label{eq:Jsis}
\end{equation}
$\tilde{\Gamma}$ has all the entries equal to $\Gamma_{11}=\lambda$ in the first row, and all others zero. This is because $\Gamma_{\nu 1}$ encodes the interactions that change the prevalence of the $\nu$-th compartment by acting with $\compart{wt}$ on the susceptible state. Hence $\nu$ itself can refer only to $\compart{wt}$, and only $wt$ is transmitted, thus the value $\lambda$. We now wish to show that $\Lambda$ is block-upper-triangular. One diagonal block, $\Lambda_1$, encompasses the indices $\beta=1,\ldots,2^v-1$, while the other, $\Lambda_2$, the remaining $\beta=2^v,2^v+1$:
\begin{equation}
 \Lambda =
 \left(
\begin{array}{c|c}
 \Lambda_1 \; [2^v-1 \times 2^v-1] & \cdots \\
 \hline
 0 & \Lambda_2 \; [2\times 2] \\
\end{array}
\right).
\end{equation}
The lower left block is clearly zero, because the transitions that change the prevalence of $\compart{seg},\compart{all}$ by acting on $\compart{wt}$ with a compartment other than $\compart{seg},\compart{all}$, or vice versa, are not possible. The block $\Lambda_1$ is upper diagonal, and we can show this with a reasoning similar to the one for $\Gamma$ in the main text. First of all, $\Lambda_{1,11}=0$ as no term $x_1^2$ exists in the equations. We then consider the transitions $\compart{\alpha}\compart{wt}\rightarrow\compart{\alpha}\compart{\beta}$, with $1< \alpha,\beta \leq 2^v-1$. It must be that $\phi(\alpha)\geq\phi(\beta)$, implying $\alpha\geq\beta$. Furthermore, the diagonal elements are $\Lambda_{1,\alpha\alpha}=\lambda^{\phi(\alpha)-1}$, as one needs to transmit all the segmented variants $\compart{\alpha}$ contains, but not $wt$. Finally, the transitions $\compart{wt}\compart{\alpha}\rightarrow\compart{wt}\compart{\beta}$ are not possible, as all the compartments considered already contain the $wt$. This proves the upper diagonal shape. The block $\Lambda_2$ does not change in dimension with $v$, and so it can be computed explicitly by analyzing the four possible reactions between $\compart{seg},\compart{all}$. Summing up, the matrices involved have the following form:
\begin{equation}
 \Gamma =
 \left(
\begin{array}{cccccc|cc}
 \lambda & \sbab        & \cdots & \sbab         & \cdots & \sbab        & \sbab               & \sbab \\ 
  0 & \lambda^2        & \cdots & \sbab         & \cdots & \sbab        & \sbab               & \sbab \\ 
 \vdots    & \vdots & \cdots & \vdots & \cdots & \vdots &\vdots        & \vdots \\ 
 0 & 0        & \cdots & \lambda^n        & \cdots & \sbab        & \sbab               & \sbab \\ 
 \vdots    & \vdots & \cdots & \vdots & \cdots & \vdots &\vdots        & \vdots \\ 
 0 & 0        & \cdots & 0         & \cdots & \lambda^{v}        & \sbab               & \sbab \\ 
 \hline
 0 & 0        & \cdots & 0         & \cdots & 0        & \lambda^v & \lambda^v (1-\lambda) \\ 
  0 & 0       & \cdots & 0         & \cdots & 0        & 0 & \lambda^{v+1} \\ 
\end{array}
\right),
\label{eq:GammaExpl}
\end{equation}
\begin{equation}
 \Lambda =
 \left(
\begin{array}{cccccc|cc}
 0 & \sbab        & \cdots & \sbab         & \cdots & \sbab        & \sbab               & \sbab \\ 
  0 & \lambda        & \cdots & \sbab         & \cdots & \sbab        & \sbab               & \sbab \\ 
 \vdots    & \vdots & \cdots & \vdots & \cdots & \vdots &\vdots        & \vdots \\ 
 0 & 0        & \cdots & \lambda^{n-1}        & \cdots & \sbab        & \sbab               & \sbab \\ 
 \vdots    & \vdots & \cdots & \vdots & \cdots & \vdots &\vdots        & \vdots \\ 
 0 & 0        & \cdots & 0         & \cdots & \lambda^{v-1}        & \sbab               & \sbab \\ 
 \hline
 0 & 0        & \cdots & 0         & \cdots & 0        & -\lambda & 0 \\ 
  0 & 0       & \cdots & 0         & \cdots & 0        & \lambda (1+\lambda^{v-1}) & \lambda^v \\ 
\end{array}
\right),
\label{eq:LambdaExpl}
\end{equation}
\begin{equation}
 \tilde{\Gamma} =
 \left(
\begin{array}{cccccc|cc}
 \lambda & \lambda        & \cdots & \lambda         & \cdots & \lambda        & \lambda               & \lambda \\ 
  0 & 0        & \cdots & 0         & \cdots & 0        & 0               & 0 \\ 
 \vdots    & \vdots & \cdots & \vdots & \cdots & \vdots &\vdots        & \vdots \\ 
 0 & 0        & \cdots & 0         & \cdots & 0        & 0               & 0 \\
 \vdots    & \vdots & \cdots & \vdots & \cdots & \vdots &\vdots        & \vdots \\ 
0 & 0        & \cdots & 0         & \cdots & 0        & 0               & 0 \\ 
 \hline
 0 & 0        & \cdots & 0         & \cdots & 0        & 0               & 0 \\ 
  0 & 0        & \cdots & 0         & \cdots & 0        & 0               & 0 \\ 
\end{array}
\right),
\label{eq:GammaTildeExpl}
\end{equation}
where $2\leq n\leq v$, and the symbol ``$\sbab$'' marks values that are not necessary to our computation. By putting Eqs.~(\ref{eq:GammaExpl}), (\ref{eq:LambdaExpl}) and~(\ref{eq:GammaTildeExpl}) into Eq.~(\ref{eq:Jsis}) we realize that $J^{(\mathbf{wt})}$ has the same block structure, and can compute its eigenvalues. The stability condition then translates into the following system:
\begin{equation}
\begin{cases}
 \mu-\lambda < 0 \\
 \lambda^n - \mu + \lambda^{n-1}\left(1-\frac{\mu}{\lambda}\right)\left( 1-\lambda \right) <0 \\
 \lambda^v - \mu <0 \\
 \lambda \left( \mu\lambda^{v-1}-1 \right) < 0 \\
\end{cases}.
\end{equation}
 The first equation is always true, as we are above $T_1$. The second one, for $n=2$, is true when $\lambda < 2\mu / (1+\mu)$. Then, if this holds, one can show that all the following hold. As a result, we get to $T_2=\left\{ \lambda = 2\mu / (1+\mu) \right\}$.

\section{Heterogeneous contacts}
\label{sec:hetdeg}

We now address the fully general equation driving the system, Eq~(\ref{eq:general}). We assume here $k$ to be discrete-valued. One can prove that the whole derivation holds in the continuous case, too.
The general form of the Jacobian Eq.~(\ref{eq:general}) becomes
\begin{align}
 J_{\nu\alpha}^{km} =  \frac{\partial  \dot{x}_\nu^k }{\partial x_\alpha^m } = & \frac{k}{\pan{k}} m p_\gamma(m) \left[  \Gamma_{\nu\alpha} \left(1-\sum_\gamma x_\gamma^k \right) + \gamma \Lambda_{\nu\alpha\gamma} x_\gamma^k \right] + \nonumber\\
 & + \delta^{km} \left[ -\mu\delta_{\nu\alpha} + \frac{k}{\pan{k}} \sum_{\beta,h} h p_\gamma (h) x_\beta^h \left( \Lambda_{\nu\beta\alpha}-\Gamma_{\nu\beta} \right) \right].
\end{align}
This matrix acts on the space $\mathcal{G}\otimes\mathcal{H}$, where $\mathcal{G}$ is the usual $(2^v+1)$-dimensional space spanned by the compartments, and $\mathcal{H}$ is an $\infty$-dimensional separable Euclidean space spanned by the discrete degrees (or contact rates). For this reason, we can study the spectrum of $J$ on the compartment sector, and the degree sector, one at the time.

\subsection*{Critical surface $T_1$}

On the \dfsi, the Jacobian reads
\begin{equation}
 \left. J_{\nu\alpha}^{km} \right|^{(\mathbf{wt})} = C_\gamma \frac{k m^{1-\gamma}}{\pan{k}} \Gamma_{\nu\alpha} - \mu \delta^{km} \delta_{\nu\alpha}.
\end{equation}
In Sect.~\ref{sec:generic} we already have examined both $\Gamma$ and $\Lambda$ thoroughly. Hence, we can say that the principal eigenvalue of $\Gamma$ is $\lambda-\mu$, corresponding to some eigenvector $v$. If one defines the vector $\kappa$ on $\mathcal{H}$ as simply the sequence of positive natural numbers $\kappa = \left(1\;2\;3\;\cdots\right)$, then one can show that $\kappa\otimes v$ is the principal eigenvector of $\left. J \right|^{(\mathbf{wt})}$, with eigenvalue $\pan{k^2}\lambda/\pan{k}-\mu$. From this we find $T_1=\left\{\lambda=\hat{\mu}\right\}$.

\subsection*{Critical surface $T_2$}

Let us call $z^k\udef x_1^k$. The {\bfseries wt} equilibrium will be some $z^k>0$, and $x_\nu^k=0$ $\forall \nu>1$. This leads to (dropping the superscript {\bfseries wt} from now on)
\begin{align}
J_{\nu\alpha}^{km} = &\frac{m p_\gamma(m)}{\pan{k}} k \left[ \Gamma_{\nu\alpha} (1-z^k) + \Lambda_{\nu\alpha 1}z^k \right] + \\
& + \delta^{km} \left[ -\mu\delta_{\nu\alpha} + k \pan{z}_{\gamma-1} \left(\Lambda_{\nu 1 \alpha}-\Gamma_{\nu 1}\right)  \right],
\end{align}
where $\pan{z}_\sigma$ is the average of $z^k$ computed with $p_\sigma (k)$: $\pan{z}_\sigma \udef C_\sigma \sum_k z^k k^{-\sigma}$. By using the findings in Sect.~\ref{sec:generic}, we know that, in the compartment sector, the relevant (dominant) eigenvalue is the entry $\nu=\alpha=2$. Hence, we can directly compute the Jacobian for these values, and deal with the degree sector:
\begin{equation}
 J^{km} = -\mu \delta^{km} + \frac{m p_\gamma(m)}{\pan{k}} k \left[ \lambda^2 (1-z^k) + \lambda z^k \right].
\end{equation}
We now define two vectors (in $\mathcal{H}$): $\Omega_k\udef k p_\gamma(k)/\pan{k}$, and $\Psi_k \udef k \left( \lambda^2 + \lambda(1-\lambda) z^k \right)$. With them we can rewrite $J^{km}$:
\begin{equation}
 J = -\mu + \Psi \Omega^T.
\end{equation}
The principal eigenvector of $J$ is $\Psi$, and the corresponding eigenvalue is $-\mu + \Omega^T \Psi$. By computing it, and setting it to zero, we recover the equation for the critical point:
\begin{equation}
 \lambda + \left(1-\lambda\right) \pan{z}_{\gamma-2} = \frac{\pan{k}}{\pan{k^2}} \frac{\mu}{\lambda}.
 \label{eq:criticalk}
\end{equation}
The last piece of the puzzle is computing the term $\pan{z}_{\gamma-2}$. We define the following function:
\begin{equation}
 g(a,x) \udef \sum_{k=1}^\infty \frac{k^{-a}}{1+x k}.
  \label{eq:g0}
\end{equation}
For this function, one can prove the following recursion relation:
\begin{equation}
 xg(a-1,x) = \zeta(a) - g(a,x)
\end{equation}
(derivation not shown here), where $\zeta$ is the Riemann zeta function. Now, from~\cite{PastorSatorras2001s} we know that the prevalence $z^k$ of a SIS model obeys the following equation:
\begin{equation}
 z^k = \frac{\lambda \pan{z}_{\gamma-1} }{ \mu + \lambda \pan{z}_{\gamma-1} }.
\end{equation}
We apply $C_{\gamma-1}\sum_k k^{1-\gamma}$ to both sides of this equations, and get
\begin{equation}
 g\left(\gamma-2, \frac{\lambda}{\mu}\pan{z}_{\gamma-1} \right) = \frac{\mu}{\lambda C_{\gamma-1}}.
 \label{eq:g1}
\end{equation}
We then apply $C_{\gamma-2}\sum_k k^{2-\gamma}$, and get
\begin{equation}
 \pan{z}_{\gamma-2} = \pan{z}_{\gamma-1} \frac{\lambda}{\mu} g\left(\gamma-3, \frac{\lambda}{\mu}\pan{z}_{\gamma-1} \right).
  \label{eq:g2}
 \end{equation}
Moreover, we notice that the moments of the degree distribution can be expressed in terms of the normalization constants as follows:
\begin{equation}
 \pan{k^n} = \frac{C_\gamma}{C_{\gamma-n}}.
  \label{eq:g3}
\end{equation}
By combining Eqs.~(\ref{eq:g0}), (\ref{eq:g1}), (\ref{eq:g2}) and~(\ref{eq:g3}), we can get to a closed-form solution for $\pan{z}_{\gamma-2}$:
\begin{equation}
 \pan{z}_{\gamma-2} = 1-\frac{\pan{k}}{\pan{k^2}} \frac{\mu}{\lambda}.
\end{equation}
Finally, by putting this inside Eq.~(\ref{eq:criticalk}), we get to $T_2 = \left\{\lambda = 2\hat{\mu} / \left(1+\hat{\mu}\right)\right\}$.

\section{Enhanced segment transmissibility}

Segmented variants now transmit with a probability $\rho\lambda$, with $\rho\geq 1$, where $\lambda$ is the transmissibility of $wt$. Let us examine how the interaction matrices change according to this. Matrix $\tilde{\Gamma}$ in Eq.~(\ref{eq:GammaTildeExpl}) does not change, while matrix $\Gamma$ in Eq.~(\ref{eq:GammaExpl}), and $\Lambda$ in Eq.~(\ref{eq:LambdaExpl}), change as follows:
\begin{equation}
 \Gamma =
 \left(
\begin{array}{cccccc|cc}
 \lambda & \sbab        & \cdots & \sbab         & \cdots & \sbab        & \sbab               & \sbab \\ 
  0 & \rho\lambda^2        & \cdots & \sbab         & \cdots & \sbab        & \sbab               & \sbab \\ 
 \vdots    & \vdots & \cdots & \vdots & \cdots & \vdots &\vdots        & \vdots \\ 
 0 & 0        & \cdots & \rho^{n-1}\lambda^n        & \cdots & \sbab        & \sbab               & \sbab \\ 
 \vdots    & \vdots & \cdots & \vdots & \cdots & \vdots &\vdots        & \vdots \\ 
 0 & 0        & \cdots & 0         & \cdots & \rho^{v-1}\lambda^{v}        & \sbab               & \sbab \\ 
 \hline
 0 & 0        & \cdots & 0         & \cdots & 0        & \left(\rho\lambda\right)^v &\left(\rho\lambda\right)^v (1-\lambda) \\ 
  0 & 0       & \cdots & 0         & \cdots & 0        & 0 & \rho^v\lambda^{v+1} \\ 
\end{array}
\right),
\label{eq:GammaExplDd}
\end{equation}
\begin{equation}
 \Lambda =
 \left(
\begin{array}{cccccc|cc}
 0 & \sbab        & \cdots & \sbab         & \cdots & \sbab        & \sbab               & \sbab \\ 
  0 & \rho\lambda        & \cdots & \sbab         & \cdots & \sbab        & \sbab               & \sbab \\ 
 \vdots    & \vdots & \cdots & \vdots & \cdots & \vdots &\vdots        & \vdots \\ 
 0 & 0        & \cdots & \left(\rho\lambda\right)^{n-1}        & \cdots & \sbab        & \sbab               & \sbab \\ 
 \vdots    & \vdots & \cdots & \vdots & \cdots & \vdots &\vdots        & \vdots \\ 
 0 & 0        & \cdots & 0         & \cdots & \left(\rho\lambda\right)^{v-1}        & \sbab               & \sbab \\ 
 \hline
 0 & 0        & \cdots & 0         & \cdots & 0        & -\lambda & 0 \\ 
  0 & 0       & \cdots & 0         & \cdots & 0        & \lambda+\left(\rho\lambda\right)^{v} & \left(\rho\lambda\right)^{v} \\ 
\end{array}
\right),
\label{eq:LambdaExplDd}
\end{equation}
The spectrum of $J^{\mbox{\scriptsize\dfsi}}=\Gamma-\mu$ then gives the first critical point. We find two regimes. For $\rho < \mu^{-\frac{v-1}{v}}$, we find the usual $T_1=\left\{\lambda=\mu\right\}$. For $\rho > \mu^{-\frac{v-1}{v}}$ a different critical point emerges: $T_s = \left\{\lambda = \left(\mu\right)^{1/v}/\rho \right\}$. This means that if transmissibility is enhanced enough, the compartment $\compart{seg}$ spreads alone as an SIS, and $T_s$ is its epidemic threshold.

In the regime $\rho < \mu^{-\frac{v-1}{v}}$ we can find the new $T_2 = \left\{ \lambda = \frac{1+\rho}{\rho} \frac{\mu}{1+\mu} \right\}$, with the same derivation as in Sect.~\ref{sec:generic}. Analogously we can add heterogeneous contact rates, solving the degree sector as in Sect.~\ref{sec:hetdeg}, finding the correction $\hat{\mu}$ as before.


\section{Total prevalence of the wild-type virus}
\label{sec:totalwtprevalence}

Total prevalence of the {\itshape wt} can be computed analytically in the homogeneous case. In order to prove that, we consider Eq.~\ref{seq:general_homo}. For convenience, we define $z = \sum_\alpha x_\alpha - x_{seg}$, which is the total prevalence of the {\itshape wt}. Primed summation symbols ($\sum_\nu'$) mean $\nu$ runs over all the compartments but $\compart{seg}$. We apply $\sum_\nu'$ to both sides of Eq.~\ref{seq:general_homo}, getting
\begin{equation}
 \dot{z} = -\mu z + \sum_{\alpha\beta} x_\alpha x_\beta \left( \sum_\nu^{}{\vphantom{\sum}}' \Lambda_{\nu\alpha\beta} \right) + (1-z-x_{seg}) \sum_\alpha x_\alpha \left( \sum_\nu^{}{\vphantom{\sum}}' \Gamma_{\nu\alpha} \right).
 \label{eq:totprevwt}
\end{equation}
The term containing $\Lambda_{\nu\alpha\beta}$ can be computed using that $\sum_\nu \Lambda_{\nu\alpha\beta} = 0$. This is due to the fact that the number of hosts is conserved, and $\Lambda_{\nu\alpha\beta}$ encodes interactions only between infected compartments. As a result, $\sum_\nu' \Lambda_{\nu\alpha\beta} = -\Lambda_{seg,\alpha\beta}$. Moreover, one can show that $\Lambda_{seg,\alpha\beta} = -\lambda \delta_{\beta,seg} (1-\delta_{\alpha,seg})$. The term $\sum_\nu' \Gamma_{\nu\alpha}$ is the probability of $\alpha$ generating a $\nu\not=seg$ by infecting a susceptible. This is just the probability of transmitting the {\itshape wt}, because all the other probabilities cancel out. Hence, $\sum_\nu' \Gamma_{\nu\alpha} = \lambda (1-\delta_{\alpha,seg})$. By inserting these two terms in Eq.~(\ref{eq:totprevwt}), one gets
\begin{equation}
 \dot{z} = -\mu z + \lambda (1-z) z,
\end{equation}
which decouples from the other variables, and represents a pure SIS. As a result, the endemic total prevalence of the {\itshape wt} in case of homogeneous networks is always $z = 1-\mu/\lambda$.


\section{Numerical validation of the critical surfaces}

\begin{figure}[htbp]
\centering
\includegraphics[width=.9\textwidth]{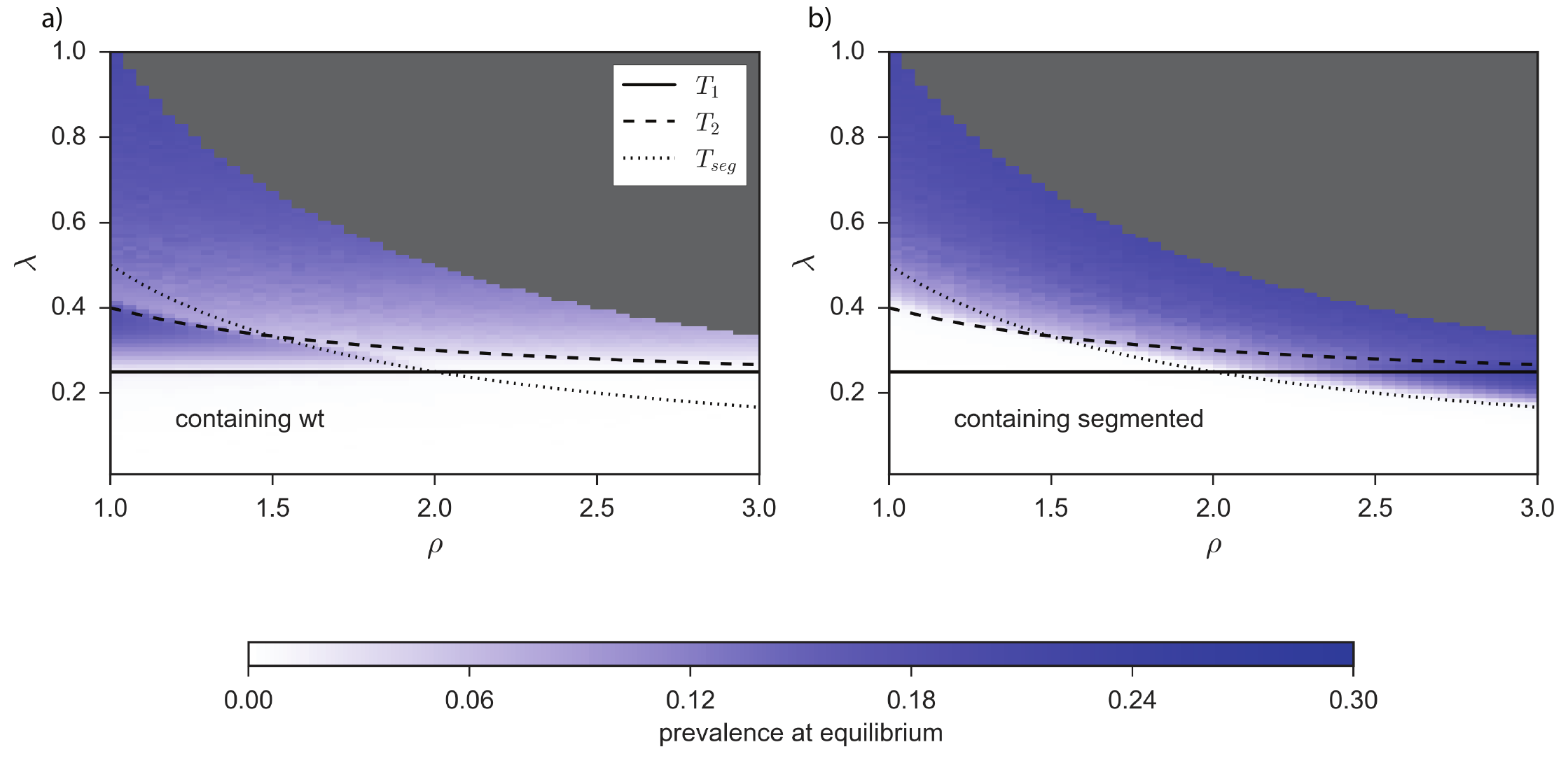} 
\caption{Numerical validation of Fig.~3(E). Prevalence of the $wt$ (A) and the segmented variants (B) are computed through stochastic simulations. The solid, dashed and dotted lines represent $T_1$, $T_2$ and $T_{s}$, respectively.}
\label{fig:panel_numeric}
\end{figure}

In order to validate our theoretical prediction of the phase space, we simulate the spread of a multipartite virus on a plant population. The estimate of the critical surfaces requires computing the endemic states, corresponding to the different phases. To do that, we used the quasistationary state method~\cite{Ferreira2011,Ferreira2012}. In its original formulation for an SIS model, the quasistationary state method relies on forcing the system out of the disease-free state. Every time the simulation produces a fully susceptible population, one inputs an active configuration previously visited by the system. With multipartite viruses, however, there is an additional challenge, represented by the fact that the disease-free state is not the only absorbing state. Every time the system becomes free of a specific variant (or $wt$) disappears from the system, it will be free of it forever. Hence, we force the system out of any state that does not contain all the $v$ variants and the $wt$. The result of the simulations is shown in Fig.~\ref{fig:panel_numeric}.

\section{Limited carrying capacity}

Our model assumes that the transmission probability of one variant does not depend on the coinfecting variants. In reality, however a limited carrying capacity should be taken into account, as the number of viral particles a cell can produce in time is limited. Here we investigate this aspect using a simple assumption: coinfecting variants share equally a fixed transmissibility. Hence, for instance, while compartment $\compart{wt}$ transmits the \wt{} with probability $\lambda$, $\compart{1}$ transmits it with probability $\lambda /2$, due to the concurrent infection by a defective variant.
We show that while this impacts the specific values of the critical surfaces in Eq. (2,3,4) of the main text, it does not change the quantitative behavior.

Having previously demonstrated the generalizability to arbitrary number of variants ($v$) and arbitrary heterogeneous topology, we set ourselves in the (computationally) simplest scenario of homogeneous network and $v=2$. Equation~(\ref{eq:v1}) thus becomes
\begin{equation}
\begin{cases}
\dot{z} = \lambda (1-z)z -\mu z - \frac{\lambda}{2} (1-z) y \\
\dot{x}_2 = \left(\frac{\lambda}{2}\right)^2 (1-z) x_2 +\frac{\lambda}{2} (z-x_2) x_2  - \mu x_2.
\end{cases}.
\label{eq:carrCap}
\end{equation}

From this, and from the SIS-like dynamic of $\compart{seg}$ spreading alone, we can compute the new critical surfaces. We consider, for simplicity, $\rho=1$ and homogeneous topology:
\begin{align}
T_1 & = \left\{\lambda =\mu \right\} ; \label{eq:T1_cc} \\
T_2 & = \left\{ \lambda = \frac{3 \mu}{1+\mu/2} \right\}; \label{eq:T2_cc} \\
T_s & = \left\{ \lambda = v \mu^{1/v} \right\}. \label{eq:Ts_cc}
\end{align}
By comparing them to Eq. (2,3,4) of the main text, we see that limited carrying capacity does not change the epidemic threshold ($T_1$). It increases, however, both $T_2, T_s$, making multipartitism overall less likely. It however does not change the qualitative behavior of the model.

\section{Nonhomogeneous recovery rates, nonindependent transmission}

Our model assumes recovery rate is the same for all compartment. One might instead assume that it either decreases or increases with the number of coinfecting variants. Here we investigate a different recovery rate for the pure multipartite compartment ($\compart{seg}$): compartment containing \wt{} recover at a rate $\mu$, $\compart{seg}$ at a rate $\sigma \mu$.

Analogously, one might assume that variants in the pure multipartite compartment do not spread independently. To that end, we introduce another correction factor $\lambda^v\longrightarrow \alpha\lambda^v$.
It is straightforward to show that both corrections impact $T_2$ (Eq.~(4)) in the same way, with the identification $\alpha = 1/\sigma$. The new critical surface containing both factor is
\begin{equation}
T_s  = \left\{ \lambda = \frac{1}{\rho} \left( \frac{\sigma \hat{\mu}}{\alpha} \right)^{1/v} \right\}.
\end{equation}
From this we see that both these assumptions add an additional scaling to the effective recovery rate $\hat{\mu} \longrightarrow \sigma \hat{\mu} / \alpha$, while leaving the overall behavior of the model unchanged.

\section*{Supporting references}

\end{document}